**Environmental Control of Triplet Emission in Donor-Bridge-Acceptor Organometallics**

*Jiale Feng, Lupeng Yang, Alexander S. Romanov, Jirawit Ratanapreechachai, Saul T. E. Jones, Antti-Pekka M. Reponen, Mikko Linnolahti, Timothy J. H. Hele, Anna Köhler, Heinz Bässler, Manfred Bochmann, Dan Credgington**

J. Feng, L. Yang, J. Ratanapreechachai, S. T. E. Jones, A. P. M. Reponen, Dr. T. J. H. Hele, Dr. D. Credgington
Cavendish Laboratory, Department of Physics, University of Cambridge, JJ Thomson Avenue, Cambridge CB3 0HE, United Kingdom.
E-mail: djnc3@cam.ac.uk

Dr. A. S. Romanov, Prof. M. Bochmann
School of Chemistry, University of East Anglia, Earlham Road, Norwich NR4 7TJ, United Kingdom.

Prof. M. Linnolahti
Department of Chemistry, University of Eastern Finland, Joensuu Campus, FI-80101 Joensuu, Finland.

Prof. A. Köhler, Prof. H. Bässler
Experimental Physics II and Bayreuth Institute of Macromolecular Science (BIMF), Department of Physics, University of Bayreuth, Bayreuth 95440, Germany.

**Abstract**

Carbene-metal-amides (CMAs) are a promising family of donor-bridge-acceptor molecular charge-transfer emitters for organic light-emitting diodes (OLEDs). Here a universal approach is introduced to tune the energy of their charge-transfer emission. A shift of up to 210 meV is achievable in the solid state via dilution in a polar host matrix. The origin of this shift has two components: constraint of thermally activated triplet diffusion, and electrostatic interactions between the guest molecules and the polar host. This allows the emission of mid-green CMA archetypes to be blue shifted without chemical modifications. Monte-Carlo simulations based on a Marcus-type transfer integral successfully reproduce the concentration- and temperature-dependent triplet diffusion process, and reveal a substantial shift in the ensemble density of states in polar hosts. In gold-bridged CMAs this substantial shift does not lead to a significant change in luminescence lifetime, thermal activation energy, reorganisation energy or intersystem crossing rate. These discoveries thus offer new experimental and theoretical insight in to the coupling between the singlet and triplet manifolds in these materials. Similar emission tuning can be achieved in related materials where chemical modification is used to modify the charge-transfer energy.

## 1. Introduction

Thin-film organic light-emitting diodes (OLEDs) have developed into a flourishing commercial industry in the last few decades. In 1987, Tang and Van Slyke demonstrated first 'sandwich structure' OLED utilising fluorescent emission from spin-singlet states.[1] Second-generation phosphorescent OLEDs utilising emission from spin-triplet states, developed a decade later, exhibit high efficiency and significant synthetic tuneability, making them the current best candidates for lighting and display technologies.[2] However, efficient deep blue OLEDs remain one of the key challenges limiting their broader application, due to low quantum efficiency and short operational lifetime.[3] Within this sphere, a new class of donor-bridge-acceptor carbene-metal-amide (CMA) emitters has been developed that exhibit high quantum efficiency at high brightness. Initial reports have centred around variants of the archetype CMA1, employing cyclic (alkyl)(amino)carbene (CAAC) acceptor, Au bridge and carbazole donor, which exhibits high photoluminescence quantum efficiency (80-90%), good chemical stability and fast intersystem crossing (~5 ps).[4] Photoluminescence arises primarily via the triplet state, and is thermally-activated, with activation energy of 70-80 meV and characteristic emission lifetime less than 1 μs in both solution-processed and thermally-evaporated architectures at 300 K.[4,5] This is shorter than efficient iridium-based phosphorescent emitters (typical lifetime > 1.5 μs )[6,7] and many emitters operating via E-type (thermally-activated) delayed fluorescence (TADF, typical lifetime > 5 μs).[8,9] Submicrosecond emission lifetime is critical for efficient high brightness OLED devices and for preventing bimolecular exciton annihilation reactions, which are implicated in reducing operational stability.[10,11] In both solution and amorphous thin film, CMA1 is a mid-green emitter. However, unlike many other triplet-harvesting organic and organometallic archetypes, CMA materials exhibit three features which allow additional routes to tune emission characteristics: 1) significant geometric flexibility, allowing tuning of excited state energies through control of geometry, 2) large negative absorption solvatochromism, indicating a large electrostatic dipole in the ground state 3) lack of concentration quenching, allowing flexibility in host:dopant ratio and host choice.

Here we show how the emission of CMA1 can be tuned by utilising intermolecular interactions with a variety of host materials to both restrict triplet diffusion and shift the density of excited states. The result is that the photoluminescence of CMA1 can be blue-shifted by 210 meV, into the blue colour range, without altering its chemical structure. We determine that electrostatic interactions are one of the most important parameters for these composites. Despite the significant change in emission energy which may be achieved, we find that the low activation energy for delayed emission and short room-temperature emission lifetimes are preserved. We use these new experimental findings to test current quantum-chemical descriptions of CMA emission, and provide a better understanding of the CMA emission mechanism. We go on to show that this approach extends to other gold CMA complexes across the visible spectrum, allowing the emission energy for each to be tuned over a 150-200 meV range.

## 2. Results and Discussion

### 2.1. Introduction to CMA1 molecule

The chemical structure of CMA1 is illustrated in **Figure 1**a. In the ground state $S_0$, the CAAC group is relatively electron deficient, while the amide is electron rich, creating a ground-state electrostatic dipole moment of order 15 D aligned along the C-Au-N axis, hereafter taken as the *z*-axis of the molecule.[12] The HOMO and LUMO of CMA1, calculated by density functional theory (DFT) utilising the hybrid MN15 functional with the def2-TZVP basis set are presented in Figure 1b**.** The HOMO resides primarily on the amide, the LUMO on the carbene. Of order 3.0/11.4% of the electron density in the HOMO/LUMO resides on the metal, with largest contributions from the $5d_{yz}/5p_y$ atomic orbitals, respectively, where the carbazole is taken to lie in the *x*-*z* plane. Excitation from $S_0$ to $S_1$ is dominated by a HOMO-LUMO transition (natural transition orbitals comprise 98% HOMO-LUMO), which spans the metal bridge and has significant charge-transfer (CT) character, shifting electron density back from the amide to the carbene group. This reduces the electrostatic dipole to approximately 5 D and reverses its sign[4,13]. In addition to direct absorption to the singlet CT state, optical absorption spectra of CMA1 show features related to ligand-centred excitations of the carbene and amide groups, Figure S1. For CMA1, the photoexcited CT singlet crosses to the triplet manifold within around 5 ps, with subsequent unstructured emission on sub-microsecond timescales.[4] The steady-state emission peak is around 520 nm (green region) in neat film at 300 K, see **Figure 2**a, with a Stokes' shift of 795 meV (measured peak-to-peak). The large Stokes' shift has been assigned in previous reports to a combination of fast vibrational relaxation followed by torsional relaxation from a coplanar to a twisted geometry, which narrows the $S_1$-$S_0$ energy gap.[13]

In solution, the CT band of CMA1 exhibits large negative absorption solvatochromism, consistent with the ordering of polar solvents around the large ground-state dipole, stabilising $S_0$ and destabilising $S_1$.[12] Emission shows weak positive solvatochromism, consistent with much weaker ordering of solvent around the smaller $S_1$ dipole, see Figure S2 and S3. In neat thin film, the CT absorption band is broader and peaks at 388 nm, between the values of toluene (407 nm) and dichloromethane (385 nm) solutions.

### 2.2 Dopant concentration and role of diffusion for CMA1 in PVK host

We first examine the effect of host on emission energy in the absence of significant electrostatic interactions. Poly(9-vinylcarbazole) (PVK) is a common polymer host material for solution processed OLEDs, possessing relatively low polarity.[14] We have previously reported efficient CMA1:PVK solution-processed OLEDs, using a CMA1 concentration of 20% by weight.[4] Figure 2a presents the steady state absorption and photoluminescence spectra of CMA1 doped in to PVK host at concentrations from 100% (neat CMA1) to 5% by weight, representing the range of dopant concentrations over which efficient OLED devices have been shown to be achievable.[5] The absorption of CMA1 in PVK is dominated by parasitic host absorption and a weak scattering tail, which obscures the exact CMA1 absorption edge. PL data are more revealing; when decreasing the concentration of CMA1 in the host-guest composite from 100% to 5%, the position of the steady-state photoluminescence peak energy blue-shifts by approximately 60 meV, from 2.39 to 2.45 eV. Photoluminescence quantum efficiency (PLQE) of these films ranges from 67% to 94%, calculated using the De Mello method.[15] The luminescence lifetime of CMA1 in host remains relatively constant, varying from 0.84 μs to 1.04 μs between 100% and 5% concentration. These values are tabulated in Table S1.

Figure 2c presents time-resolved PL peak energy over time as a function of concentration at room temperature. Spectral migration is observed over the lifetime of the excited state, with migration rate dependent on dopant concentration. The spectral relaxation shifts on a logarithmic time scale with increasing concentration. This implies that the diffusion process takes place via the dopant, and via an electronic coupling between dopants that depends exponentially on distance, such as a Dexter-type transfer.[16] At high dopant concentration, the peak position saturates to a constant value at long times. At low dopant concentration, migration is too slow for saturation to be observed before PL becomes undetectable. Migration rates for both high and low concentration are reduced at low temperature, shown in Figure 2e and f and steady state PL peak energies blue shift with decreasing temperature. The luminescence rate of CMA1 in PVK is strongly thermally activated above 120 K, increasing by nearly two orders of magnitude between 10 K and 300 K, with characteristic activation energy of 72 meV for 10% CMA1 and 76 meV for 80% CMA1, see Figure 2d and Figure S4. The total time-integrated luminescence increases with the same activation energy Figure S5, indicating that thermal activation is primarily of the radiative triplet decay rate. Calculations from thermally activated decay rates are shown in Equation S1 and Equation S2. The observed temperature dependence is consistent with the diffusion of triplet excitons, for example as reported in neat films of poly(p-phenylene) type conjugated polymers.[17]

We model this behaviour by considering that spectral migration occurs via triplet diffusion through a disordered density of emitter states, as described by Movaghar et al.[18] By applying a Monte-Carlo simulation of 3D triplet diffusion, we find that a Dexter-type dependency of hopping probability on intermolecular distance reproduces the observed concentration-dependent migration rate, with the long-time saturation of peak position occurring where triplets are able to relax to the tail of their density of states. We find that a fixed density of states for all concentrations is sufficient to model the trend observed. Moreover, we find that a Marcus-type activated hopping probability (Equation S3) is required to reproduce the observed temperature dependence.[19] By fitting the experimental data, we extract a characteristic reorganisation energy $\lambda$ of 240 meV, corresponding to an activation energy of $E_a = \lambda/4$ = 60 meV. We note that the trend observed cannot be reproduced using a Miller-Abrahams type hopping probability, which does not account for reorganisation.[17] Emission is in general from a non-equilibrium ensemble of triplet excited states, with restriction of triplet diffusion able to tune emission energy over a small (60 meV) range. Only at high dopant concentration and higher temperatures are photoexcited triplets able to relax to a quasi-equilibrium energy in the tail of the density of states within their emission lifetime.[20] Note that the steady state emission energy reflects the weighted time-integrated signal so that it contains contributions from across the DOS. At quasi-equilibrium in solid films, PL peak energies are within ~10 meV of those observed in low polarity solvents (benzene, toluene), which we take as evidence that the tail of the density of states represents molecules close to the fully-relaxed $S_1$ geometry. The blue shift in steady-state PL at decreased temperature results from the reduction in diffusion rate outcompeting the decrease in emission rate.

To explore the generality of this effect, **Figure 3** presents the dependence of steady-state luminescence peak energy on CMA1 concentration for a range of polymer and small molecule host materials with high triplet energies, deposited as thin films from solution. For all hosts lacking a significant permanent electric dipole, calculated by DFT (B3LYP/6-31G**), a universal blue-shift is observed as concentration decreases, very close in magnitude to that observed in PVK. We interpret this as evidence that triplet diffusion between guest emitters is primarily limited by emitter spacing, and relatively insensitive to the nature of the intervening host. We likewise infer that molecular relaxation on the timescales of triplet emission plays only a minor role.

By contrast, we observe an additional shift in host molecules exhibiting permanent electric dipole moments, specifically Bis(N-carbazolyl)benzene (mCP) with 1.4 D; Bis[3,5-di(9H-carbazol-9-yl)phenyl]diphenylsilane (SimCP2) with 2.37 D; 9-(3-(9H-Carbazol-9-yl)phenyl)-3-(diphenylphosphoryl)-9H-carbazole (mCPPO1) with 3.91 D and Diphenyl-4-triphenylsilylphenyl-phosphine oxide (TSPO1) with 4.1 D.

Chemical structures, electric dipole moments and steady state photoluminescence spectra of these hosts are shown in Figure S6 and S7. Steady state luminescence energy increases markedly compared to non-polar host matrices for mCP and TSPO1, with the trend of PL in mCP host consistent with that observed in electroluminescence by Conaghan et al.[5] The effect for larger host molecules SimCP2 and mCPPO1 is less pronounced. To understand the origin of this phenomenon, we focus on TSPO1, which produces the largest magnitude shift in this set.

### 2.3 Organic polar molecule hosts and role of electrostatic interactions

Steady-state absorption and photoluminescence spectra of CMA1 in TSPO1 host at various concentrations are shown in **Figure 4**a. As for PVK, the absorption of CMA1 in TSPO1 is dominated by parasitic host absorption and a weak scattering tail, which renders the exact absorption edge difficult to resolve. However, photoluminescence peak position shows a large blue shift compared to nonpolar hosts, by approximately 210 meV when decreasing the weight-concentration of CMA1, from 2.39 eV for neat CMA1 films to 2.6 eV for 5:95 wt.% CMA1:TSPO1 films. PLQE of these films are around 65% to 80%. Luminescence lifetime increases slightly from 0.97 μs (neat film) to 1.4 μs (5% CMA1), see Table S2. Low-temperature luminescence lifetime of 10% CMA1 in TSPO1 increases by approximately a factor of 50, from 1.3 μs at 300 K to 65 μs at 10 K, see Figure 4d. The same trend is seen at higher concentration, luminescence lifetime increases from 1.02 μs at 300 K to 66 μs at 10 K for 80% CMA1 in TSPO1, Figure S8 The activation energies extracted from PL decay rate of 10% and 80% CMA1 in TSPO1 host are 79 meV and 77 meV, which are close to the values for PVK hosted samples. Integrated PL intensity against temperature of 10% CMA1 in TSPO1 also yields the same activation energy as extracted from PL decay rate, indicating that thermal activation still contributes primarily to the radiative triplet decay rate as shown in Figure S9.

Monte-Carlo modelling of concentration- and temperature-resolved emission spectra Figure 4c, e, f reveal that this shift has two components. The first is a thermally activated spectral migration, consistent with triplet diffusion via an activated hopping process, with characteristic reorganisation energy $\lambda = 240$ meV, the same as that observed for CMA1:PVK films. However, we are unable to reproduce the trend observed assuming a fixed density of states. Instead, the mean energy of the Gaussian density of states varies with concentration, shifting up by 113 meV between neat film and 5% concentration at 300 K. At the same time, the deviation narrows by 25 meV. By examining the low-temperature spectral diffusion for both high and low concentration, see Figure 4e and f, it is clear that this effect becomes more pronounced at low temperature, shifting the PL spectra to higher energy.

XRD and GIWAXS measurements indicate there is very little evidence of crystallisation of CMA1 in these TSPO1 composites Figure S10 and S11, though at low CMA1 concentration, weak host crystallisation features are observed. We therefore propose that the energetic shift is due to an electrostatic interaction between the large ground state electric dipole moment of CMA1 with the smaller host dipoles, leading to orientation of the latter during deposition. Previous work by Dos Santos et al. also showed that polar matrix is able to rearrange molecular configurations and shifts the energy of CT state.[21] The effect is increased at low temperature, which we interpret as a reduction in thermal disorder. This solid-state solvatochromism leads to a stabilisation of the ground state, and a destabilisation of the excited state, as is observed in liquid solution. Unlike in solution, upon excitation the host dipoles are less able to reorient, preserving the increased energetic splitting between the ground and excited states. As for low-polarity hosts, this is consistent with diffusion, rather than intramolecular relaxation, dominating spectral relaxation in the solid state. Consistent with this, we find no correlation with measured glass transition temperatures (which are > 100 ˚C for all hosts) or with molecular weight, see Figure S12. An additional effect is the apparent narrowing of the density of states in the MC model, which requires some explanation. We consider that this is most likely an effect of the microstructure of low-concentration TSPO1 blends. The appearance of a weak TSPO1 crystallisation signal suggests an inhomogeneous local microstructure, which might reduce access to the full DOS and manifest as such a narrowing.

We thus conclude that of the 210 meV shift in steady-state luminescence observed between neat and dilute CMA1, approximately 60 meV arises from supressed triplet diffusion and 150 meV arises from an electrostatic host-guest interaction. In addition to providing a mechanism by which emission peak energy may be controlled over a meaningful range, solid-state solvatochromism offers a means to tune the relative energies of excited states with differing charge-transfer character, and probe the underlying mechanism of triplet harvesting.

Initial TD-DFT calculations of CMA1 using the PBE0 functional and def2-TZVP basis set, when referenced to experimental energies, suggested torsion leads to a crossing of the lowest singlet and triplet energies.[4] However, TD-DFT calculations using the more accurate MN15 functional, which does not suffer from the underestimation typical for TD-DFT reveal a different picture, predicting a significantly greater destabilisation of $S_0$ by torsion, and a reduced stabilisation of $S_1$, see Figure S13, such that both the $S_1 – S_0$ and $T_1 – S_0$ energy gaps reduce as dihedral angle increases.[22] This is consistent with the combined DFT and multireference configuration interaction calculations of Föller and Marian, who find that while the $S_1$ state is stabilised to a greater extent than $T_1$ by contributions from doubly-excited configurations, this is insufficient in most circumstances to invert the spin states.[12] This work also concluded that $T_1$ phosphorescence borrows oscillator strength from $S_2$, and that coupling between $S_1$ and $T_1$ is spin-vibronic in nature. Subsequent work by Penfold et al. considered couplings from $S_1$ to the lowest three excited triplet states and two nuclear degrees of freedom: torsion around the Au–N bond and the stretching mode of the same bond with the molecule in the relaxed $S_1$ geometry.[23] This work concluded that indirect SOC (i.e. $S_1 – T_n –$

$T_1$) mediated by torsional motion may influence the rate of triplet harvesting. Taffet et al. in examining the structurally related Cu(I) analogue CMA2 concluded that ISC was likely most effective in a sterically constrained coplanar configuration, relying on a breaking of planar symmetry by distortion of the C-Cu-N central axis to allow coupling between $S_1$ and $T_1$.[24] The underlying process coupling the $T_1$ state to the singlet manifold is therefore unclear. We use the experimental results above to provide new insight to this question. While charge-transfer excited states show significant negative absorption solvatochromism, excited states localised to the donor and acceptor ligands ("LE" states) are insensitive to environmental polarizability, see Figure S2 and S3. Upon dilution in TSPO1 host, the peak-to-peak energy difference between the lowest-lying emissive triplet localised to the carbazole donor (2.9 eV) (Figure S14) and the CT triplet decreases from 510 meV to 300 meV. A similar analysis considering the shift in high-energy edge gives a range 190 meV to around 0 meV. Despite being brought substantially closer to resonance with the localised triplet, increased CT energy leads to a slight *increase* in emission lifetime, and no meaningful change in activation energy is observed. We also find that excited state lifetime depends very weakly on whether or not a given excited molecule is structurally relaxed, with lifetime in dilute solid comparable to lifetime in dilute solution, see Table S3. Intersystem crossing (ISC) time is likewise insensitive to the increased CT energy, and is constant at around 5 to 6 ps, see Figure S15.

To establish a framework for interpreting these observations, we consider a simplified version of the chromophore which has $C_{2v}$ symmetry[25], with reference to molecular orbital calculations summarised in figures S16-18 and Table S10. From an examination of symmetry arguments, the locally excited Carbazole state (≈ HOMO to LUMO+3) transforms as $A_1$ in the $C_{2v}$ point group, the same irrep as the $S_1$ and $T_1$ states. In $C_{2v}$ the molecular rotations (and therefore the SOC operator) transform as all irreps except $A_1$. When the CMA chromophore is completely planar or twisted at 90 degrees (both of which correspond to approximate $C_{2v}$ symmetry), there can therefore be no direct SOC between $S_1$ and $T_1$, and nor can there be any indirect SOC via the Cz LE state (referred to as $LE_1$).

Since there can be no direct coupling (and therefore RISC) in this idealised picture, further coupling has to be included in the model. One possibility is mixing via higher-lying states which transform as irreps other than $A_1$. The ligand-centred state corresponding to ≈ HOMO-3 to LUMO (referred to as $LE_2$) transforms as $B_2$ and therefore its triplet form can mix with $S_1$. Similarly, the state formed by a predominantly HOMO-1 to LUMO transition is predominantly of CT character (referred to as $CT_2$) and transforms as $B_1$, thus its triplet form can interact with $S_1$ via SOC. In order for $LE_2$ and $CT_2$ to mix with $T_1$ there would need to be vibronic symmetry breaking, for $LE_2$ a $B_2$ mode and for $CT_2$ a $B_1$ mode. It is impossible to state for certain without calculation which of these interactions is stronger or more likely to contribute to SOC and therefore delayed fluorescence. However, the insensitivity of the photophysics to the energy of the CT states relative to the LE states suggests that the interaction could be via $CT_2$, which will be perturbed by electrostatic environment similarly to $S_1$. In addition, upon descending from $C_{2v}$ (the symmetry of the model chromophore) to $C_s$ (the symmetry of the actual chromophore) $B_1$ descends to A'; as does $A_1$, whereas $A_2$ and $B_2$ descend to A''. This suggests that in the reduced symmetry of the true chromophore then $CT_2$ may be more able to couple with $T_1$ than $LE_2$. While these arguments consider interactions via the triplet forms of $LE_2$ and $CT_2$, similar arguments also hold for interactions via the singlet forms, though this is in reality less likely due to larger energy separation. An alternative mechanism for coupling to occur between states of CT character could be vibronically allowed SOC, particularly since calculation suggests emission may result from a configuration involving twisting around the carbene-metal-amide bond.[26] In a partially twisted geometry (between 0 and 90 degrees) the idealised chromophore descends in symmetry to $C_2$ (and the full chromophore to $C_1$), and $S_1$ and $T_1$ descend from $A_1$ to A. In this lower symmetry $R_z$ also transforms as A, meaning that in the twisted geometry there could be direct SOC between $S_1$ and $T_1$, facilitating emission. This would also be consistent with the relative insensitivity in the rates of SOC with host polarizability, since the $S_1 - T_1$ energy gap is likely to be similar in range of host environments.

### 2.4 Extension to other CMAs

This strategy for emission tuning and photophysical exploration does not rely on a particular emitter choice, the same approach can be extended to tune the emission of other charge-transfer emitters with a large ground-state dipole moment. In particular, we apply the same approach to emitters from the CMA family across the visible spectrum. **Figure 6**a presents thin-film photoluminescence peak energies for CMA1 and structurally-related gold-bridged analogues: (CAAC)Au(3,6-di-tBucarbazole) (CMA4), (CAAC)Au(6-(tert-butyl)-3-(trifluoromethyl)-9H-carbazole) (CMA5), (CAAC)Au(10,11-dihydrodibenz[b,f]azepin-5-ide) (CMA6) embedded in the TSPO1 host. We observe a universal blue shift of PL peak energies as concentration is reduced, with similar magnitude to that of CMA1, around 200 meV.

## 3. Conclusion

In summary, we demonstrated a physical approach to modulate the CT triplet energy in a donor-bridge-acceptor type organometallic emitter CMA1. The CT energy can be tuned by around 200 meV via thermally activated diffusion and electrostatic interactions with host molecules. This shift leads to no meaningful change in intersystem crossing rate, slightly increased luminescence lifetime, and no significant change in thermal activation energy. The energy relaxation process was studied by Monte-Carlo simulations, which show that triplet diffusion can account for the experimental trends if modelled as a Marcus-type rate equation and the mean energy and width of the density of states responds to the variations of host-guest electrostatic interactions. We infer that structural relaxation is hindered in the solid state as we see little evidence for large-amplitude structural relaxation or host reorganisation occurring during the excited state lifetime. Tuning of CT energy provides an experimental approach to probe the triplet harvesting mechanism and the coupling between the $T_1$ and $S_1$ states. We find that for CMA1 there is likely to be no direct spin-orbit coupling between charge transfer states ($S_1$ and $T_1$) and ligand-centred excited states localised to the carbazole, as they transform as the same irrep. However, higher-lying CT states (for example LUMO-1 to HOMO) can interact with $S_1$ and $T_1$. From the insensitivity of CMA1 photophysics to CT energy, we suggest that CT – CT coupling contributes more significantly than CT – LE coupling, offering a design rule for the realisation of rapid triplet emission. We go on to show that solid-state solvatochromism may be applied to a range of gold-bridged CMAs, achieving a universal blue shift around 200 meV. Such an approach should be directly transferrable to other emitters with large permanent dipole moment, allowing host-guest interactions as a tool to tune electroluminescence in OLED devices over a significant range.

## 4. Experimental Section

### *4.1 Synthesis of carbene metal amides*

*General Considerations.* Unless stated otherwise, all reactions were carried out in air. Solvents were distilled and dried as required. Sodium *tert*-butoxide, 3-(*tert*-butyl)phenylboronic acid, were purchased from FluoroChem, SPhos Pd G2 was purchased from Sigma-Aldrich and used as received. The carbene ligand ($^{Ad}$L),[27–29] *N*-(2-chloro-4-(trifluoromethyl)phenyl)acetamide and 6-(*tert*-butyl)-3-(trifluoromethyl)-9H-carbazole carbazole,[30] and complex ($^{Ad}$L)AuCl[31] were obtained according to literature procedures. $^{1}$H and $^{13}$C{$^{1}$H} NMR spectra were recorded using a Bruker Avance DPX-300 MHz NMR spectrometer. $^{1}$H NMR spectra (300.13 MHz) and $^{13}$C{$^{1}$H} (75.47 MHz) were referenced to $CD_2Cl_2$ at δ 5.32 ($^{13}$C, δ 53.84), $C_6D_6$ at δ 7.16 ($^{13}$C, δ 128.4), $CDCl_3$ at δ 7.26 (δ $^{13}$C 77.16) ppm. Elemental analyses were performed by London Metropolitan University.

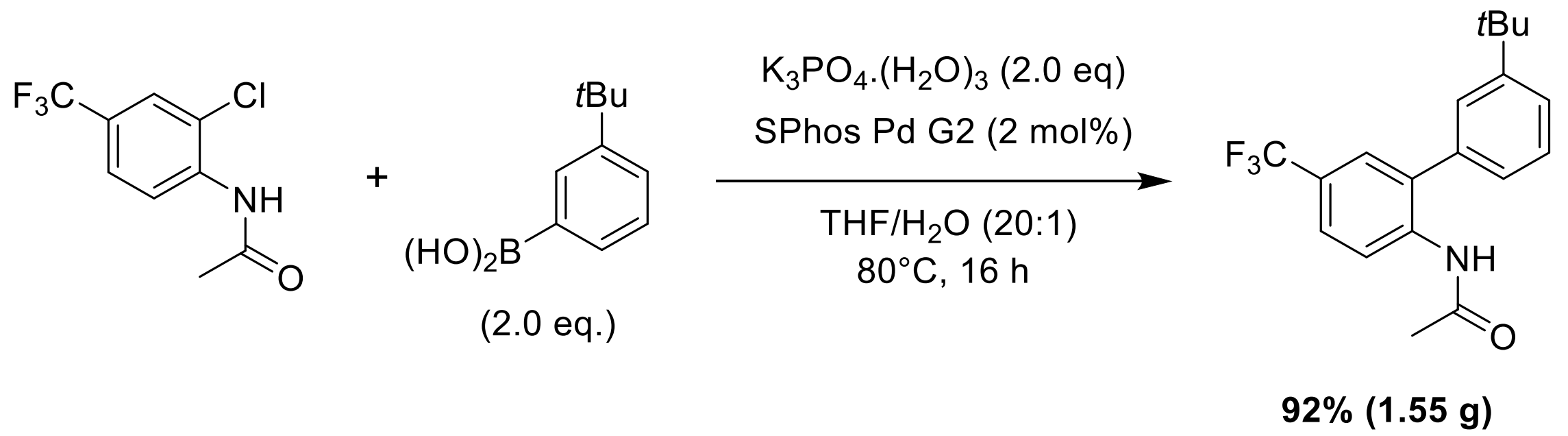

*Synthesis of N-(3'-(tert-butyl)-5-(trifluoromethyl)-[1,1'-biphenyl]-2-yl)acetamide.* *N*-(2-chloro-4-(trifluoromethyl)phenyl)acetamide (1 eq., 5.05 mmol, 1.20 g), 3-(*tert*-butyl)phenylboronic acid (2.0 eq., 10.1 mmol, 1.80 g,) and potassium phosphate trihydrate (2.0 eq., 10.1 mmol, 2.33 g) were mixed in THF/$H_2O$ 20:1 (10 mL) and purged with argon. SPhos Pd G2 (1 mol%, 0.051 mmol, 37 mg) was added and the mixture was heated at 80°C for 16 h. Reaction was cooled to r.t., $Et_2O$ (30 mL) was added and the mixture was filtered through Celite®. The filtrate was diluted with AcOEt (100 mL), washed with water and brine, and dried with $MgSO_4$. The solvent was evaporated and the residue was purified by silica column chromatography (PE/AcOEt) to afford the product as an off-white solid (92%, 1.55 g).

$^1$H NMR (300 MHz, $CDCl_3$): δ 8.53 (d, *J* = 8.7 Hz, 1H), 7.61 (pseudo dd, *J* = 8.7, 2.2 Hz, 1H), 7.52 – 7.42 (m, 3 x 1H overlapped), 7.40 – 7.37 (m, 1H), 7.37 – 7.32 (bs, NH), 7.19 (pseudo dt, *J* = 6.6, 1.9 Hz, 1H), 2.05 (s, 3H, Ac), 1.37 (s, 9H, *t*Bu). $^{13}$C NMR (75 MHz, $CDCl_3$) δ 168.4 (s, C=O), 152.7 (s, C–*t*Bu), 138.0 (s, $C_q$), 136.4 (s, $C_q$), 132.2 (s, $C_q$)), 129.5 (s, CH), 127.1 (q, *J* = 3.6 Hz, CH–C–$CF_3$), 126.4 (s, CH), 126.3 (s, CH), 125.8 (s, CH), 125.5 (q, *J* = 3.7 Hz, CH–C–$CF_3$), 124.2 (q, *J* = 271.8 Hz, $CF_3$), 120.8 (s, CH), 35.1 (s, $C(CH_3)_3$), 31.5 (s, $C(CH_3)_3$), 25.0 (s, $CH_3$ Ac), ($C_{ipso}$–$CF_3$ was not observed due to overlap with aromatic signals). $^{19}$F NMR (282 MHz, $CDCl_3$) δ -62.1 ppm. Anal. Calcd. for $C_{19}H_{20}F_3NO$ 335.37): C, 68.05; H, 6.01; N, 4.18. Found: C, 68.17; H, 6.18; N, 4.31.

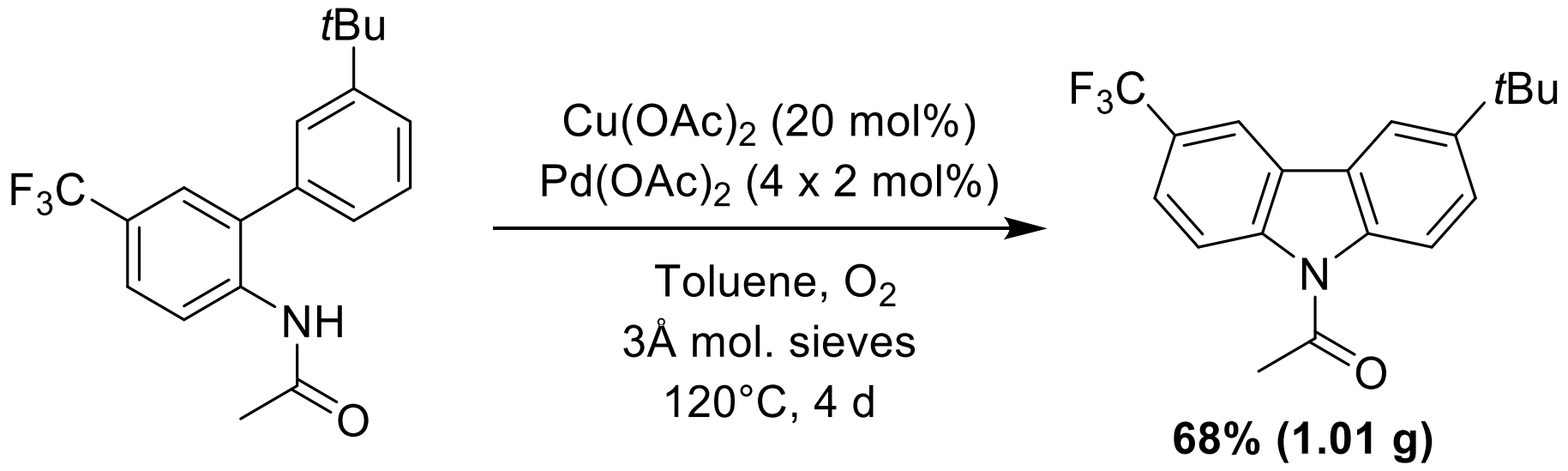

*Synthesis of 6-(tert-butyl)-3-(trifluoromethyl)-9-acetylcarbazole.* In an oven-dried Schlenk tube, under argon, *N*-(3'-(*tert*-butyl)-5-(trifluoromethyl)-[1,1'-biphenyl]-2-yl)acetamide (1 eq., 4.47 mmol, 1.50 g), $Cu(OAc)_2$ (20 mol%, 0.89 mmol, 162 mg), $Pd(OAc)_2$ (2 mol%, 0.089 mmol, 20 mg) and 3 Å molecular sieves were mixed in toluene (20 mL). The flask was purged by oxygen and heated at 120°C. Reaction completion was followed and a portion of $Pd(OAc)_2$ was added each day (3 x 2 mol%, 3 x 20 mg). After 2 days, reaction was completed. The mixture was cooled to r.t., diluted with AcOEt (60 mL) and filtered through Celite®. The filtrate was diluted with AcOEt (100 mL) washed with water and brine and dried with $MgSO_4$. The solvent was evaporated and the residue was purified by silica column chromatography (PE/AcOEt) to afford the product as an off-white solid (68%, 1.01 g).

$^{1}$H NMR (300 MHz, $CDCl_3$): δ 8.46 (d, *J* = 8.8 Hz, 1H, $CH^1$), 8.27 (pseudo s, 1H, $CH^4$), 8.04 (d, *J* = 2.1 Hz, 1H, $CH^5$), 8.01 (d, *J* = 8.9 Hz, 1H, $CH^8$), 7.71 (pseudo dd, *J* = 8.8, 1.9 Hz, 1H, $CH^2$), 7.60 (dd, *J* = 8.9, 2.1 Hz, 1H, $CH^7$), 2.90 (s, 3H, Ac), 1.45 (s, 9H, *t*Bu). $^{13}$C NMR (75 MHz, $CDCl_3$) δ 170.1 (s, C=O), 147.5 (<u>C</u>–*t*Bu), 140.9 (s, $C_q$), 137.1 (s, $C_q$), 126.8 (s, $C_q$), 126.1 (s, $CH^7$), 126.0 (q, *J* = 32.6 Hz, <u>C</u>–$CF_3$), 125.6 (s, $C_q$), 124.7 (q, *J* = 271.7 Hz, $CF_3$), 124.2 (q, *J* = 3.6 Hz, $CH^2$), 117.0 (s, $CH^1$ overlapped with $CH^4$), 116.9 (q, *J* = 3.9 Hz, $CH^4$ overlapped with $CH^1$), 116.8 (s, $CH^5$), 115.6 (s, $CH^8$), 34.9 (s, <u>C</u>($CH_3$)$_3$, 31.8 (C(<u>C</u>$H_3$)$_3$, 27.8 (s, $CH_3$ Ac). $^{19}$F NMR (282 MHz, $CDCl_3$) δ -61.2. Anal. Calcd. for $C_{19}H_{18}F_3NO$ (333.35): C, 68.46; H, 5.44; N, 4.20. Found: C, 68.13; H, 5.68; N, 3.97.

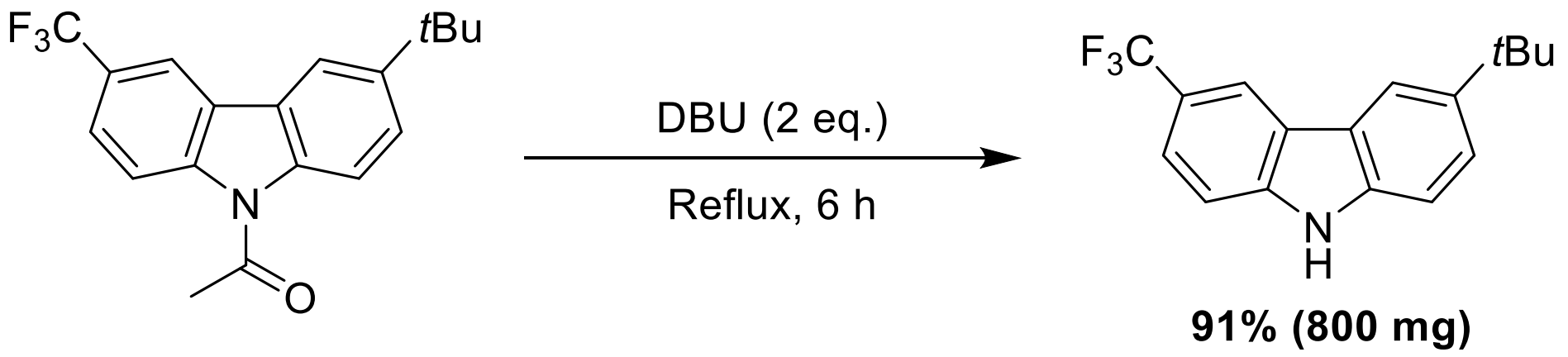

*Synthesis of 6-(tert-butyl)-3-(trifluoromethyl)-9H-carbazole.* DBU (2 eq., 6.06 mmol, 904 µL) was added to 6-(*tert*-butyl)-3-(trifluoromethyl)-9-acetylcarbazole (1 eq., 3.03 mmol, 1.01 g) in MeOH (30 mL), and the mixture was refluxed for 6 h. The reaction was cooled to r.t. and volatiles were evaporated. AcOEt (120 mL) was added, washed with water and brine, and dried over $MgSO_4$. The residue was purified by silica column chromatography (PE/AcOEt) to afford the product as a white solid (91%, 800 mg).

$^1$H NMR (300 MHz, $CDCl_3$): δ 8.37 (s, 1H, $CH^4$), 8.16 (bs, 1H, NH), 8.12 (pseudo s, 1H, $CH^5$), 7.64 (pseudo dd, $J$ = 8.5, 1.7 Hz, 1H, $CH^2$), 7.56 (dd, $J$ = 8.6, 1.9 Hz, 1H, $CH^8$), 7.47 (d, $J$ = 8.5 Hz, 1H, $CH^1$), 7.41 (d, $J$ = 8.6 Hz, 1H, $CH^7$), 1.45 (s, 9H, *t*Bu). $^{13}$C NMR (75 MHz, $CDCl_3$) δ 143.6 (s, <u>C</u>–*t*Bu), 141.5 (s, $C_q$), 138.2 (s, $C_q$), 125.5 (q, $J$ = 271.2 Hz, $CF_3$), 125.0 (s, $CH^7$), 123.4 (s, $C_q$), 122.8 (s, $C_q$), 122.5 (q, $J$ = 3.7 Hz, $CH^2$), 121.7 (q, $J$ = 32.1 Hz, <u>C</u>–$CF_3$), 117.9 (q, $J$ = 4.1 Hz, $CH^4$), 116.8 (s, $CH^5$), 110.7 (s, $CH^1$), 110.6 (s, $CH^8$), 34.9 (s, <u>C</u>($CH_3$)$_3$), 32.1 (s, C(<u>C</u>$H_3$)$_3$). $^{19}$F NMR (282 MHz, $CDCl_3$) δ -60.1. Anal. Calcd. for $C_{17}H_{16}F_3N$ (291.32): C, 70.09; H, 5.54; N, 4.81. Found: C, 69.82; H, 5.72; N, 4.63.

*Synthesis of ($^{Ad}$CAAC)Au(6-(tert-butyl)-3-(trifluoromethyl)-9H-carbazole) (CMA5).* In a Schlenk tube, ($^{Ad}$L)AuCl (3.52 g, 5.77 mmol,), 6-(*tert*-butyl)-3-(trifluoromethyl)-9H-carbazole (1.68 g, 5.77 mmol), and $^t$BuONa (0.56 g, 5.83 mmol,) were stirred in THF (75 mL) for 6 h. The mixture was filtered through Celite. The filtrate was concentrated and washed with hexane to afford the product as a white solid. Yield: 93% (4.65 g, 5.37 mmol).

$^1$H NMR (300 MHz, $CD_2Cl_2$): δ 8.20 (s, 1H, $CH^4$ Cz), 7.99 (d, $J$ = 2.0 Hz, 1H $CH^5$ Cz), 7.71 (t, $J$ = 7.8 Hz, 1H, *p*-CH Dipp), 7.47 (d, $J$ = 7.8 Hz, 2H, *m*-CH Dipp), 7.30 (dd, $J$ = 8.6, 2.0 Hz,

1H, $CH^7$ Cz), 7.25 (dd, *J* = 8.6, 1.5 Hz 1H, $CH^2$ Cz), 6.87 (d, *J* = 8.6 Hz, 1H, $CH^8$ Cz), 6.42 (d, *J* = 8.6 Hz, 1H, $CH^1$ Cz), 4.31 (d, *J* = 12.9 Hz, 2H, $CH_2$ Adamantyl), 2.90 (Sept, *J* = 6.7 Hz, 2H, CH *i*Pr Dipp), 2.45 (pseudo s, 2H + 1H, $CH_2$ CAAC overlapping with CH Adamantyl), 2.19 – 1.86 (m, 11H, Adamantyl), 1.44 (s, 6H, $C(CH_3)_2$ CAAC), 1.39 (s, 9H, *t*Bu), 1.36 – 1.30 (m, 12H, $CH_3$ *i*Pr Dipp). $^{13}C$ NMR (75 MHz, $CD_2Cl_2$) δ 244.1 (s, C: CAAC), 151.9 (s, $C_q$ Cz), 148.8 (s, $C_q$ Cz), 146.2 (s, *o*-C Dipp), 140.4 (s, C–*t*Bu), 136.7 (s, *i*-C Dipp), 130.0 (s, *p*-CH Dipp), 26.8 (q, *J* = 270.3 Hz, $CF_3$), 123.9 (s, $C_q$ Cz), 123.8 (s, $C_q$ Cz), 123.0 (s, $CH^7$ Cz), 119.8 (q, *J* = 3.2 Hz, $CH^2$ Cz), 117.1 (q, *J* = 31.3 Hz, C–$CF_3$ overlapping with $CH^4$ Cz), 116.9 (q, *J* = 4.2 Hz, $CH^4$ Cz overlapping with C–$CF_3$), 115.9 (s, $CH^5$ Cz), 114.0 (s, $CH^1$ Cz), 113.9 (s, $CH^8$ Cz), 77.6 (s, s, $C(CH_3)_2$ CAAC), 64.5 (s, C–C: CAAC), 49.1 (s, $CH_2$ CAAC), 39.4 (s, Adamantyl), 37.6 (s, Adamantyl), 35.8 (s, Adamantyl), 34.8 (s, 2 C overlapped, $C(CH_3)_3$ and CH Adamantyl) 32.2 (s, $C(CH_3)_3$) 29.6 m, 2 C overlapped, $C(CH_3)_2$ and CH *i*Pr Dipp), 28.6 (s, Adamantyl), 27.7 (s, Adamantyl), 26.5 (s, $CH_3$ *i*Pr Dipp), 23.4 (s, $CH_3$ *i*Pr Dipp). $^{19}F$ NMR (282 MHz, $CD_2Cl_2$) δ -59.1. Anal. Calcd. for $C_{44}H_{54}AuF_3N_2$ (864.89): C, 61.10; H, 6.29; N, 3.24. Found: C, 61.35; H, 6.07; N, 3.43.

*Synthesis of ($^{Ad}$CAAC)Au(10,11-dihydrodibenz[b,f]azepin-5-ide) (CMA6).* Following the procedure described for CMA5, the complex was made from ($^{Ad}$CAAC)AuCl (0.2 g, 0.33 mmol), NaO$^t$Bu (33 mg, 0.33 mmol) and 10,11-dihydro-5H-dibenz[b,f]azepine (64.3 mg, 0.33 mmol) as an orange powder. Yield: 0.169 g (0.22 mmol, 66 %).

$^1H$ NMR (300 MHz, $CD_2Cl_2$): δ 7.53 (t, *J* = 7.6 Hz, 1H, aryl), 7.30 (d, *J* = 7.6 Hz, 2H, aryl), 6.75 (d, *J* = 7.4 Hz, 2H, azepine $CH^4$), 6.69 (d, *J* = 7.4 Hz, 2H, azepine $CH^1$), 6.62 (t, *J* = 7.4 Hz, 2H, azepine $CH^3$), 6.43 (t, *J* = 7.4 Hz, 2H, azepine $CH^2$), 3.95 (d, *J* = 13.2 Hz, 2H, $CH_2$), 2.89 (s, 4H, $CH_2$ azepine), 2.83 (sept, *J* = 6.6 Hz, 2H, C*H*$Me_2$), 2.14–1.69 (m, 14H, adamantyl CH and $CH_2$), 1.36 (s, 6H, $C(CH_3)$), 1.34 (d, *J* = 6.6 Hz, 6H, CH*Me*$_2$), 1.30 (d, *J* = 6.6 Hz, 6H, CH*Me*$_2$). $^{13}C$ NMR (75 MHz, $CD_2Cl_2$) δ 242.3 (C carbene), 153.9 (*ipso*-CN azepine), 145.4

(*o*-C), 136.4 (*ipso*-C), 129.8 (*p*-CH), 129.5 (azepine $CH^4$), 127.0 (azepine *ipso*-C), 125.7 (azepine $CH^3$), 125.4 (*m*-CH), 124.5 (azepine $CH^1$), 116.7 (azepine $CH^2$), 76.6 ($C_q$), 64.1 ($C_q$), 48.8 ($CH_2$), 39.3 ($CH_2$), 37.2 (CH), 36.4 ($CH_2$), 35.0 ($CH_2$), 34.8 (azepine $CH_2$), 29.3, 28.0, 27.6, 26.0, 23.4 ($CH_3$). Anal. Calcd. for $C_{41}H_{51}N_2Au$ (768.82): C, 64.05; H, 6.69; N, 3.64. Found: C, 64.27; H, 6.83; N, 3.51.

Synthesis of *($^{Ad}$CAAC)Au(Carbazole) (CMA1)* and *($^{Ad}$CAAC)Au(3,6-di-tBucarbazole) (CMA4)* can be referred to the previous paper by Di et al.[4]

*4.2 Sample preparation*

Host-guest thin films of different weight ratios were made from chlorobenzene solution in 20 mg/mL concentration. These well-mixed solutions were spun inside a nitrogen filled glove box onto pre-cleaned spectrosil substrates at 1,200 r.p.m. for 40 s at room temperature to form thin films. Samples were stored in nitrogen glovebox to minimise degradation. Solution samples of various solvents were prepared as 1mg/ml in the nitrogen glovebox, deoxygenized and sealed in 1 mm path length and QS grade quartz cuvettes.

*4.3 UV-Vis-NIR spectrophotometer*

Shimadzu UV-3600 Plus spectrophotometer was used to measure the steady-state absorbance of samples, which comprises three detectors: a PMT (photomultiplier tube) for the ultraviolet and visible regions and InGaAs and cooled PbS detectors for the near-infrared region. The detectable wavelength range is between 185 to 3,300 nm with high resolution of 0.1 nm.

*4.4 Photoluminescence spectrometer*

The FLS980 spectrofluorimeter was used to measure steady-state luminescence spectra. A R928P PMT detector was used in this experiment, with a wavelength range of 200 nm to 870 nm and a dark count rate of <50 cps (at -20°C). The detector is operated in single photon counting mode. The PL spectra of CMA1 were collected from 350 nm to 650 nm with the resolution of 1 nm. Samples were excited by a 450 W Xe1 xenon arc lamp. The light from the xenon arc is focused into the monochromators by using an off-axis ellipsoidal mirror.

*4.5 Room temperature and cryogenic ns-μs time-resolved photoluminescence measurements*

Time-resolved photoluminescence spectra were measured by an electrically-gated intensified CCD (ICCD) camera (Andor iStar DH740 CCI-010) connected to a calibrated grating spectrometer (Andor SR303i). Samples were photoexcited by femtosecond laser pulses which were created by second harmonic generation (SHG) in a ß-barium borate crystal from the fundamental source (wavelength = 800 nm, pulse width = 80 fs) of a Ti: Sapphire laser system (Spectra-Physics Solstice), at a repetition rate of 1kHz. The photons from the laser pulses had a wavelength of 400 nm. A 425 nm long-pass filter was used to prevent scattered laser signal from entering the camera. Temporal evolution of the PL emission was recorded by stepping the ICCD gate delay with respect to the trigger pulse. The minimum gate width of the ICCD was 5 ns. Cryogenic measurements were carried out using an Oxford Instruments Optistat Dynamic continuous flow cryostat with liquid helium coolant, and an ITC 502 S temperature control unit.

The 1kHz repetition rate of the laser used in this experiment makes the maximum accurate measure of lifetimes within 1 ms. For non-exponential luminescence decays in the solid state, a characteristic lifetime rather than monoexponential decay time is quoted. We choose the time taken for the delayed component to reach 63% (1-(1/e)) of the total emission integrated from 0 to 550 μs. This allows direct comparison to lifetimes extracted from monoexponential decays as this is the fraction of total emission which has been emitted after one time constant of a monoexponential decay.

*4.6 Monte-Carlo simulation procedure*

A cubic lattice with 101x101x101 triplet sites, where site energies were drawn randomly from a Gaussian distribution, was employed for the Monte-Carlo simulation. The initial triplet excitation was placed in the centre of the lattice. The hopping rates to the nearest 125 neighbours were calculated using Equation S3. The hopping probability to each site and the overall hopping time were computed with Equation S4 and Equation S5. The triplet excitation was allowed to jump to one of the 125 sites randomly according to the hopping probabilities and the hopping time was added to the total simulation time. The procedures above were repeated until a pre-defined diffusion time was reached. The parameterisation for Figure 2 and Figure 4 is described in detail in the supplementary text.

*4.7 Transient Absorption (TA) Spectroscopy*

Films were drop-cast from solution, 60 µl per film on 13 mm quartz discs heated to 70 °C. Solutions of CMA1 and TSPO1 in chlorobenzene were made at 20 mg/ml concentrations, overnight heated at 70 °C to dissolve the TSPO1 and mixed in the appropriate ratio to achieve the desired host-guest concentration (5 wt.%). Films were made and kept in a glovebox and encapsulated using a glass slide and epoxy prior to being measured. Drop-cast films had an optical density of 0.12 at 400 nm (the pump wavelength). The main laser used was a Spectra Physics Solstice Ti:Sapphire laser, outputting pulses of width 80 fs and a repetition rate of 1 kHz at 800 nm. The pump beam was frequency-doubled using a BBO to give 400 nm pulses. Excitation fluence was varied from 9-60 $uJcm^{-2}$. The probe beam was generated from the 800 nm fundamental using a noncollinear optical parametric amplifier, built in-house. The probe was further split into a probe and reference, with only the probe beam overlapping with the pump on the sample. The pump-probe delay was controlled using a computer-controlled delay stage. A Hamamatsu G11608-512 InGaAs dual-line array detector was used to measure the transmitted probe and reference.

*4.8 Computational details*

Density functional theory (DFT) calculations of CMA1 were carried out by the global hybrid MN15 functional by Truhlar and coworkers[32] in combination with the def2-TZVP basis set by Ahlrichs and coworkers.[33,34] Relativistic effective core potential of 60 electrons was used to describe the core electrons of Au.[35] The ground state was studied by DFT and the excited states by time-dependent DFT (TD-DFT).[36] The employed method provides excited state energies that do not suffer from underestimation typical for TD-DFT,[37,38] as indicated by our recent work on closely related molecules[22,39] as well as by comparison to $T_1$ energies calculated by unrestricted DFT: The unrestricted and TD-DFT $T_1$ energies differed by only 0.004 eV. All calculations were carried out by Gaussian 16.[40] DFT calculations of dipole moments were carried out using the function B3LYP[41] with the 6-31G** basis set, and were carried out in ORCA 4.1.0[42].

**Supporting Information**

Supporting Information is available from the Wiley Online Library or from the author. The data underpinning this investigation is available at [insert URL]

**Acknowledgements**

The authors thank Russel Holmes for fruitful discussions. The authors thank the Diamond Light Source for access to the I07 beamline and for the help during the GIWAX measurements. The authors thank Edoardo Ruggeri for helpful discussions on GIWAX data. The authors thank Dr. Emrys Evans for helpful guidance on the DFT calculations of host dipole moment. J.F. acknowledges his parents for financial support on his Ph.D. L.Y. acknowledges Trinity-Barlow Scholarship. D.C acknowledges the support from the Royal Society (grant no. UF130278). S.T.E.J. acknowledges support from the Royal Society (grant no. RG140472). A.P.M.R. acknowledges support from the Royal Society (grant no. RGF\EA\180041) and the Osk, Huttunen fund. M.B acknowledges the ERC Advanced Investigator Award (grant no. 338944-GOCAT). A.S.R acknowledges support from the Royal Society (grant no. URF\R1\180288 and RGF\EA\181008). This work was supported by the EPSRC Cambridge NanoDTC, EP/L015978/1. T.J.H.H. acknowledges a Research Fellowship from Jesus College, Cambridge. A.K and H.B acknowledge financial support by the EC through the Horizon 2020 Marie Sklodowska-Curie ITN project TADF life. M.L. acknowledges the Academy of Finland Flagship Programme, Photonics Research and Innovation (PREIN), decision 320166. The computations were made possible by use of the Finnish Grid and Cloud Infrastructure resources (urn:nbn:fi:research-infras-2016072533).

Received: ((will be filled in by the editorial staff))
Revised: ((will be filled in by the editorial staff))
Published online: ((will be filled in by the editorial staff))

**References**

[1] Tang, C. W. & Vanslyke, S. A. Organic electroluminescent diodes. *Appl. Phys. Lett.* **51**, 913–915 (1987).

[2] Baldo, M. A., O 'Brien, D. F., You, Y., Shoustikov, A., Sibley, S., Thompson, M. E., Forrest, S. R., Baldo, M. A., O'Brien, D. F., You, Y., Shoustikov, A., Sibley, S. & Thompson, M. E. Highly efficient phosphorescent emission from organic electroluminescent devices. *Nature* **395**, 151–154 (1998).

[3] Lee, J., Chen, H.-F., Batagoda, T., Coburn, C., Djurovich, P. I., Thompson, M. E. & Forrest, S. R. Deep blue phosphorescent organic light-emitting diodes with very high brightness and efficiency. *Nat. Mater.* **15**, 1–8 (2015).

[4] Di, D., Romanov, A. S., Yang, L., Richter, J. M., Rivett, J. P. H., Jones, S., Thomas, T. H., Jalebi, M. A., Friend, R. H., Linnolahti, M., Bochmann, M. & Credgington, D. High-performance light-emitting diodes based on carbene-metal-amides. *Science (80-. ).* (2017).

[5] Conaghan, P. J., Menke, S. M., Romanov, A. S., Jones, S. T. E., Pearson, A. J., Evans, E. W., Bochmann, M., Greenham, N. C. & Credgington, D. Efficient Vacuum-Processed Light-Emitting Diodes Based on Carbene–Metal–Amides. *Adv. Mater.* **30**, 1–5 (2018).

[6] Baldo, M. A., Lamansky, S., Burrows, P. E., Thompson, M. E. & Forrest, S. R. Very high-efficiency green organic light-emitting devices based on electrophosphorescence. *Appl. Phys. Lett.* **75**, 4–6 (1999).

[7] Sajoto, T., Djurovich, P. I., Tamayo, A. B., Oxgaard, J., Goddard, W. A. & Thompson, M. E. Temperature dependence of blue phosphorescent cyclometalated Ir(III) complexes. *J. Am. Chem. Soc.* **131**, 9813–9822 (2009).

[8] Uoyama, H., Goushi, K., Shizu, K., Nomura, H. & Adachi, C. Highly efficient organic light-emitting diodes from delayed fluorescence. *Nature* **492**, 234–8 (2012).

[9] Tao, Y., Yuan, K., Chen, T., Xu, P., Li, H., Chen, R., Zheng, C., Zhang, L. & Huang, W. Thermally activated delayed fluorescence materials towards the breakthrough of organoelectronics. *Adv. Mater.* **26**, 7931–7958 (2014).

[10] Baldo, M. A., Adachi, C. & Forrest, S. R. Transient analysis of organic electrophosphorescence. II. Transient analysis of triplet-triplet annihilation. *Phys. Rev. B* **62**, 10967–10977 (2000).

[11] Reineke, S., Walzer, K. & Leo, K. Triplet-exciton quenching in organic phosphorescent light-emitting diodes with Ir-based emitters. *Phys. Rev. B - Condens. Matter Mater. Phys.* **75**, 1–13 (2007).

[12] Föller, J. & Marian, C. M. Rotationally Assisted Spin-State Inversion in Carbene-Mental-Amides Is an Artifact. *J. Phys. Chem. Lett.* 5643–5647 (2017).

[13] Hall, C. R., Romanov, A. S., Bochmann, M. & Meech, S. R. Ultrafast Structure and Dynamics in the Thermally Activated Delayed Fluorescence of a Carbene-Metal-Amide. *J. Phys. Chem. Lett.* **9**, 5873–5876 (2018).
[14] Yook, K. S. & Lee, J. Y. Small molecule host materials for solution processed phosphorescent organic light-emitting diodes. *Advanced Materials* (2014).
[15] John C. de, M., H. Felix, W. & H. Friend, R. An Improved Experimental Determination of External Photoluminescence Quantum Efficiency. *Adv. Mater.* **9**, 230–232 (1997).
[16] Lange, J., Ries, B. & Bässler, H. Diffusion and relaxation of triplet excitations in binary organic glasses. *Chem. Phys.* **128**, 47–58 (1988).
[17] Hoffmann, S. T., Athanasopoulos, S., Beljonne, D., Bässler, H. & Köhler, A. How do triplets and charges move in disordered organic semiconductors? A Monte Carlo study comprising the equilibrium and nonequilibrium regime. *J. Phys. Chem. C* **116**, 16371–16383 (2012).
[18] Movaghar, B., Grünewald, M., Ries, B., Bässler, H. & Würtz, D. Diffusion and relaxation of energy in disordered organic and inorganic materials. *Phys. Rev. B* **33**, 5545–5554 (1986).
[19] Köhler, A. & Bässler, H. What controls triplet exciton transfer in organic semiconductors? *J. Mater. Chem.* **21**, 4003–4011 (2011).
[20] Köhler, A. & Bässler, H. Electronic Processes in Organic Semiconductors-An Introduction. *Wiley-VCH* (2015).
[21] Dos Santos, P. L., Ward, J. S., Batsanov, A. S., Bryce, M. R. & Monkman, A. P. Optical and Polarity Control of Donor-Acceptor Conformation and Their Charge-Transfer States in Thermally Activated Delayed-Fluorescence Molecules. *J. Phys. Chem. C* **121**, 16462–16469 (2017).
[22] Romanov, A. S., Jones, S. T. E., Yang, L., Conaghan, P. J., Di, D., Linnolahti, M., Credgington, D. & Bochmann, M. Mononuclear Silver Complexes for Efficient Solution and Vacuum-Processed OLEDs. *Avdanced Opt. Mater.* **1801347**, 1–5 (2018).
[23] Thompson, S., Eng, J. & Penfold, T. J. The intersystem crossing of a cyclic (alkyl)(amino) carbene gold (i) complex. *J. Chem. Phys.* **149**, (2018).
[24] Taffet, E. J., Olivier, Y., Lam, F., Beljonne, D. & Scholes, G. D. Carbene − Metal − Amide Bond Deformation, Rather Than Ligand Rotation, Drives Delayed Fluorescence. *J. Phys. Chem. Lett.* 1620–1626 (2018).
[25] Hele, T. J. H. & Credgington, D. Theoretical description of Carbene-Metal-Amides. (2018).
[26] Evans, E. W., Olivier, Y., Puttisong, Y., Myers, W. K., Hele, T. J. H., Menke, S. M., Thomas, T. H., Credgington, D., Beljonne, D., Friend, R. H. & Greenham, N. C. Vibrationally Assisted Intersystem Crossing in Benchmark Thermally Activated Delayed Fluorescence Molecules. *J. Phys. Chem. Lett.* **9**, 4053–4058 (2018).
[27] Lavallo, V., Canac, Y., Präsang, C., Donnadieu, B. & Bertrand, G. Stable cyclic (alkyl)(amino)carbenes as rigid or flexible, bulky, electron-rich ligands for transition-metal catalysts: A quaternary carbon atom makes the difference. *Angew. Chemie - Int. Ed.* **44**, 5705–5709 (2005).
[28] Jazzar, R., Dewhurst, R. D., Bourg, J. B., Donnadieu, B., Canac, Y. & Bertrand, G. Intramolecular 'hydroiminiumation' of alkenes: Application to the synthesis of conjugate acids of Cyclic Alkyl Amino Carbenes (CAACs). *Angew. Chemie - Int. Ed.* **46**, 2899–2902 (2007).
[29] Jazzar, R., Bourg, J. B., Dewhurst, R. D., Donnadieu, B. & Bertrand, G. Intramolecular 'hydroiminiumation and -amidiniumation' of alkenes: A convenient, flexible, and scalable route to cyclic iminium and imidazolinium salts. *J. Org. Chem.* **72**, 3492–3499 (2007).

[30] Gantenbein, M., Hellstern, M., Le Pleux, L., Neuburger, M. & Mayor, M. New 4,4'-Bis(9-carbazolyl)-biphenyl derivatives with locked carbazole-biphenyl junctions: High-triplet state energy materials. *Chem. Mater.* **27**, 1772–1779 (2015).

[31] Romanov, A. S. & Bochmann, M. Gold(I) and Gold(III) Complexes of Cyclic (Alkyl)(amino)carbenes. *Organometallics* **34**, 2439–2454 (2015).

[32] Yu, H. S., He, X., Li, S. L. & Truhlar, D. G. MN15: A Kohn-Sham global-hybrid exchange-correlation density functional with broad accuracy for multi-reference and single-reference systems and noncovalent interactions. *Chem. Sci.* **7**, 5032–5051 (2016).

[33] Weigend, F. & Ahlrichs, R. Balanced basis sets of split valence, triple zeta valence and quadruple zeta valence quality for H to Rn: Design and assessment of accuracy. *Phys. Chem. Chem. Phys.* **7**, 3297–3305 (2005).

[34] Weigend, F., Häser, M., Patzelt, H. & Ahlrichs, R. RI-MP2: optimized auxiliary basis sets and demonstration of efficiency. *Chem. Phys. Lett.* **294**, 143–152 (1998).

[35] Andrae, D., Häußermann, U., Dolg, M., Stoll, H. & Preuß, H. Energy-adjustedab initio pseudopotentials for the second and third row transition elements. *Theor. Chim. Acta* **77**, 123–141 (1990).

[36] Furche, F., Rappoport, D. & Olivuccim, M. *Density functional methods for excited states: equilibrium structure and electronic spectra in Computational Photochemistry*. (Elesvier, Amsterdam, 2005).

[37] Dreuw, A. & Head-Gordon, M. Single-Reference ab Initio Methods for the Calculation of Excited States of Large Molecules. *Chem. Rev.* **105**, 4009–4037 (2005).

[38] Moore, B., Sun, H., Govind, N., Kowalski, K. & Autschbach, J. Charge-Transfer Versus Charge-Transfer-Like Excitations Revisited. *J. Chem. Theory Comput.* **11**, 3305–3320 (2015).

[39] Romanov, A. S., Yang, L., Jones, S. T. E., Di, D., Morley, O. J., Drummond, B. H., Reponen, A. P. M., Linnolahti, M., Credgington, D. & Bochmann, M. Dendritic Carbene Metal Carbazole Complexes as Photoemitters for Fully Solution-Processed OLEDs. *Chem. Mater.* **31**, 3613–3623 (2019).

[40] Frisch, M. J., Trucks, G. W., Schlegel, H. B., Scuseria, G. E., Robb, M. A. & R., C. J. Gaussian 16, Revision A. 03. Gaussian. Inc., Wallingford CT (2016).

[41] Becke, A. D. Density-functional thermochemistry. III. The role of exact exchange. *J. Chem. Phys.* **98**, 5648–5652 (1993).

[42] Neese, F. The ORCA program system. *Wiley Interdiscip. Rev. Comput. Mol. Sci.* **2**, 73–78 (2012).

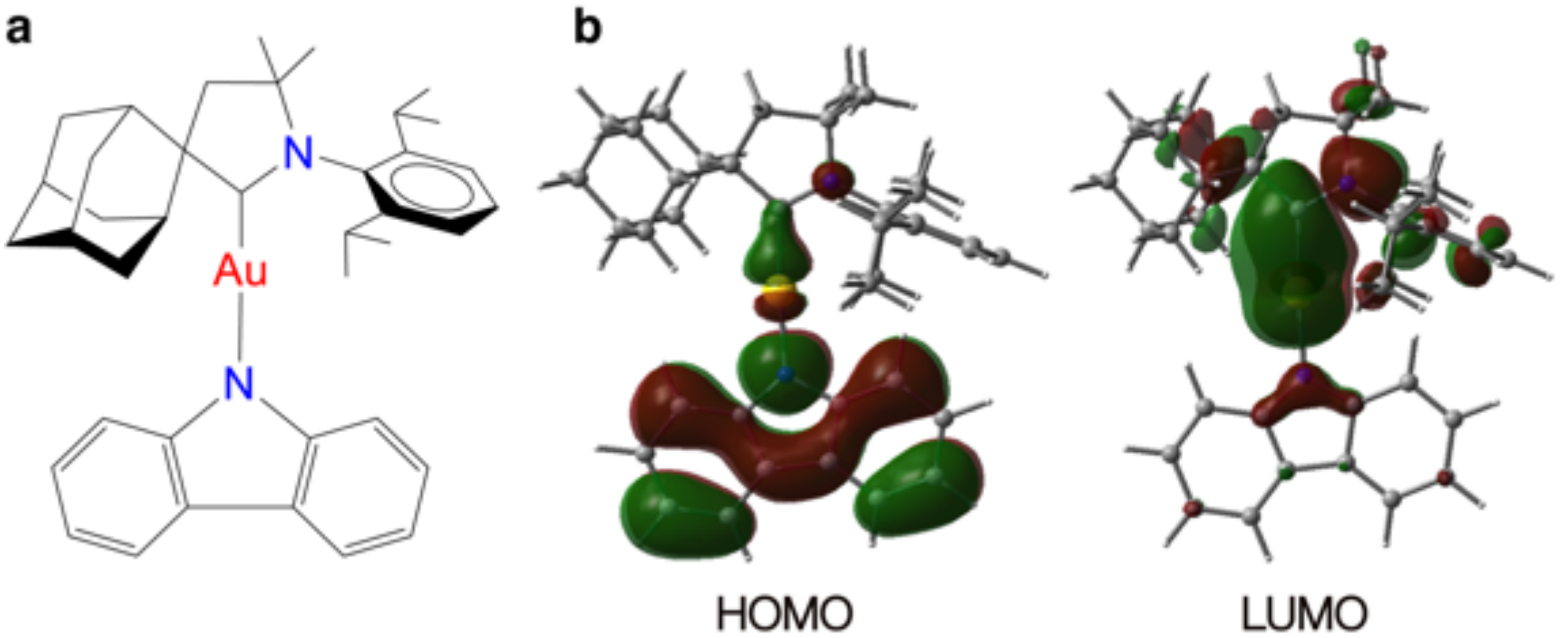

**Figure 1. a** Chemical structure of CMA1. **b** HOMO and LUMO wavefunctions of CMA1 from DFT calculations, red/green corresponds to positive/negative sign of wavefunctions.

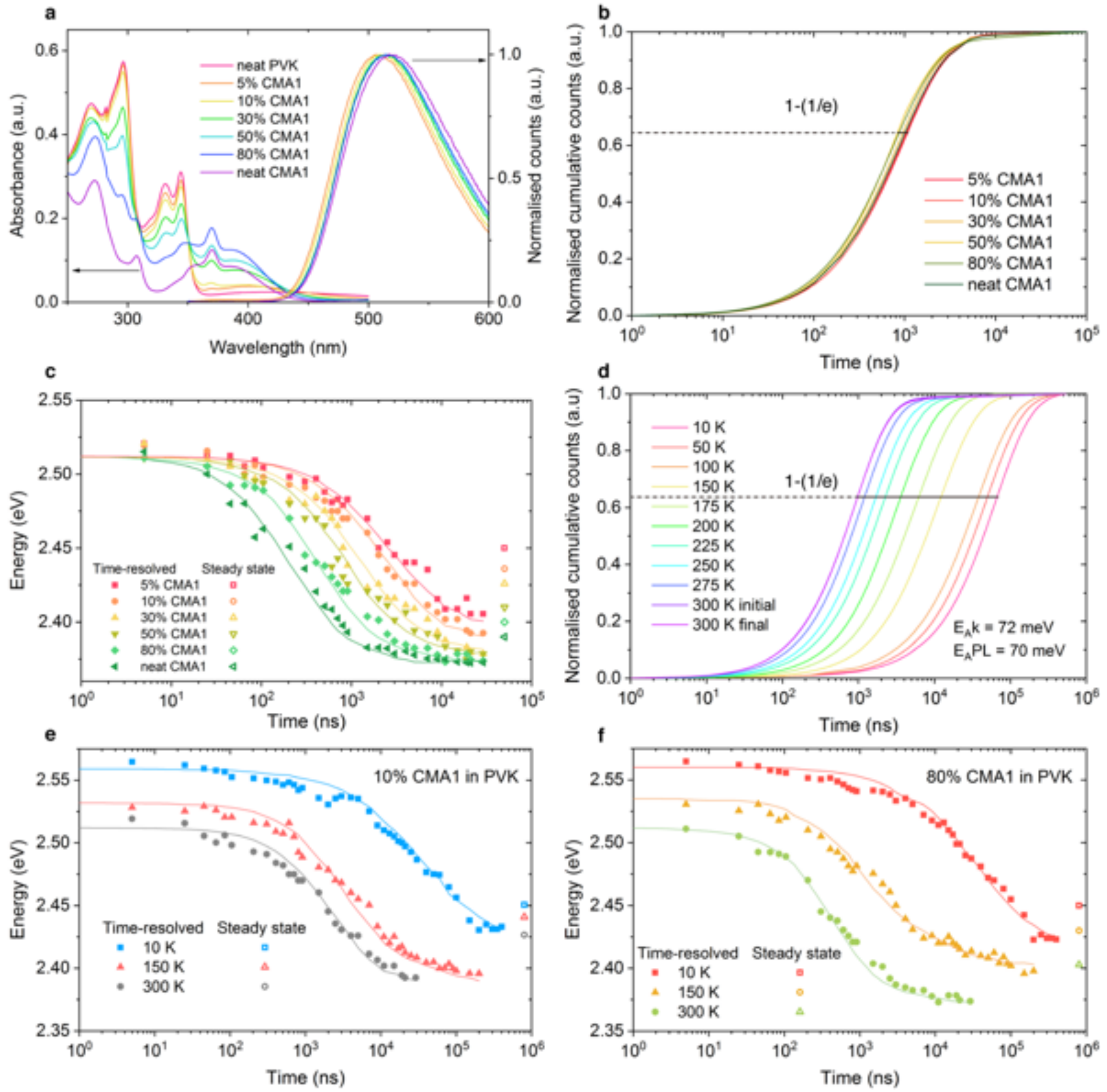

**Figure 2. a** Steady state absorption and photoluminescence of CMA1 in PVK host at different concentrations. **b** Room temperature emission integral of CMA1 and PVK composites, with 1-(1/e) labelled as the characteristic luminescence lifetime. **c** Room temperature time resolved PL peak energy of CMA1 in PVK at different concentrations to track the spectral diffusion. Lines are results of Monte-Carlo simulation. **d** Cryogenic emission integral of 10% CMA1 in PVK with 1-(1/e) labelled as the characteristic luminescence lifetime. "Initial data taken at 300 K before cooling the film to 10 K, "Final" data upon warming back to 300 K after low-temperature measurements. Two activation energies are labelled, $E_Ak$ is extracted from the PL decay rate against temperature, and $E_APL$ is extracted from the integrated PL intensity against temperature. **e**, **f** Cryogenic time resolved PL peak position of 10% and 80% CMA1 in PVK at 10 K, 150 K, and 300 K. Lines are results of Monte-Carlo simulation.

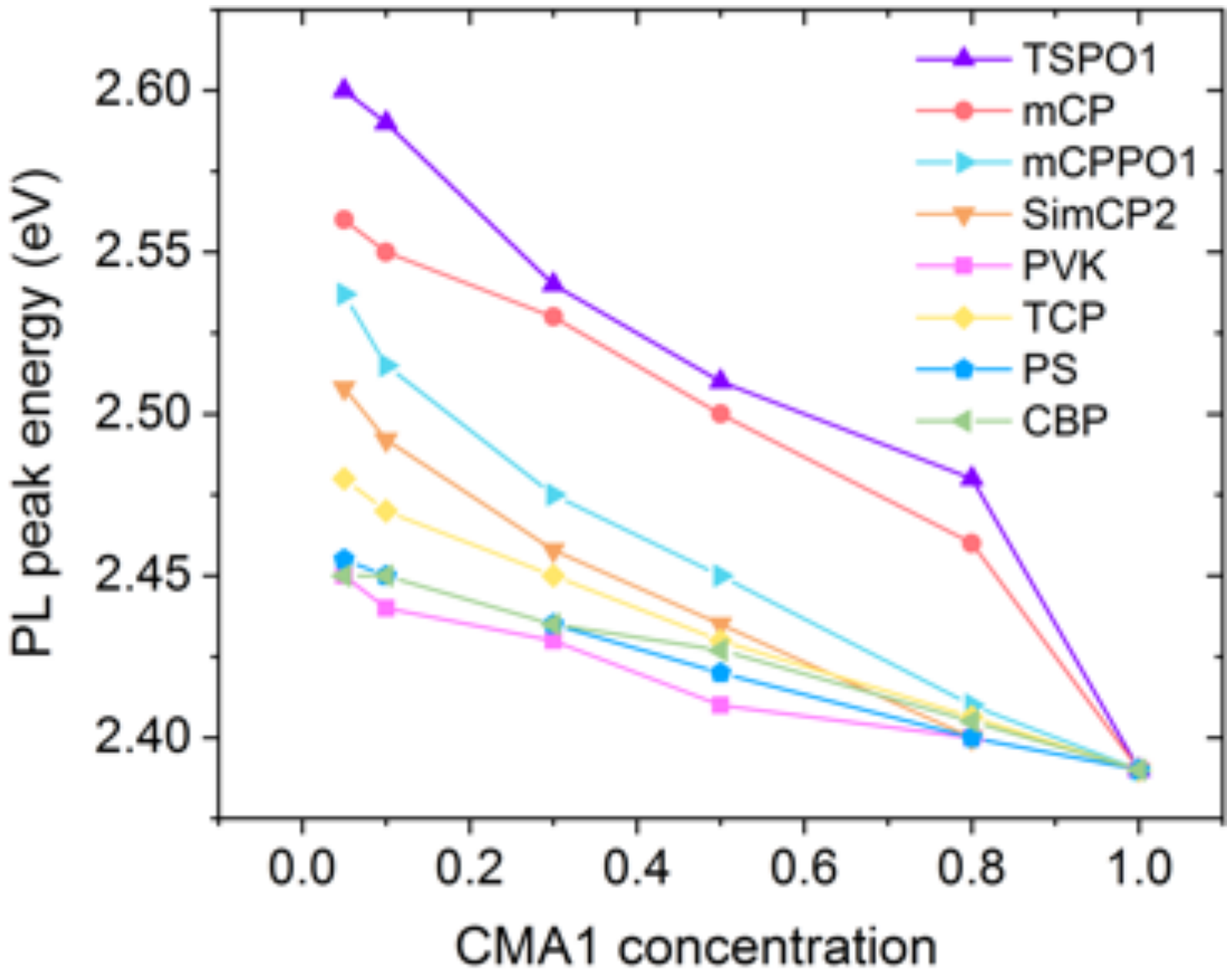

**Figure 3:** Steady state PL peak energy of CMA1 in a range of hosts varying weight concentrations from 100% to 5%.

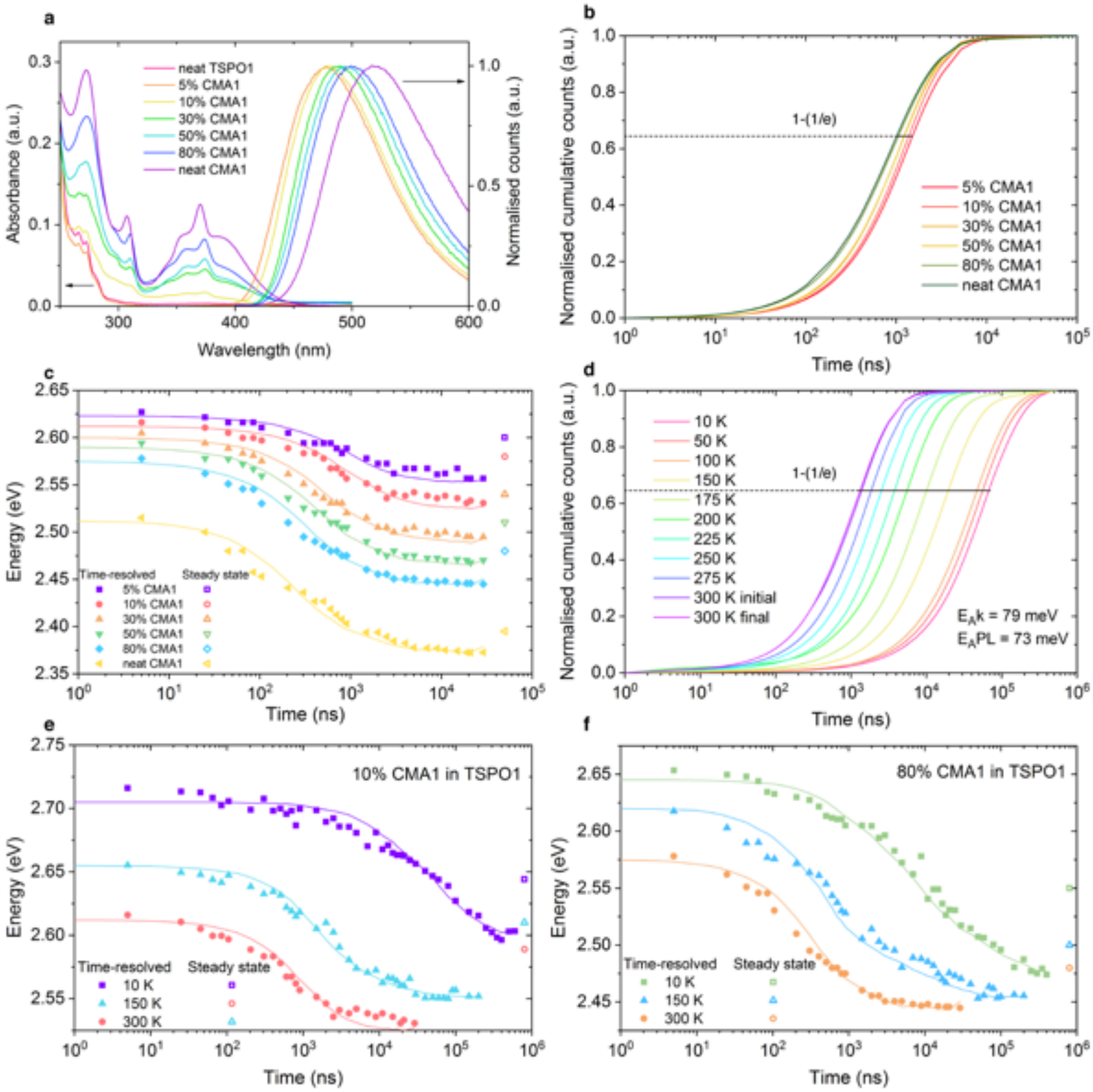

**Figure 4. a** Steady state absorption and photoluminescence of CMA1 in TSPO1 host at different concentrations. **b** Room temperature emission integral of CMA1 and TSPO1 composites, with 1-(1/e) labelled as the characteristic luminescence lifetime. **c** Room temperature time resolved PL peak energy of CMA1 in TSPO1 at different concentrations to track the spectral diffusion. Lines are results of Monte-Carlo simulation. **d** Cryogenic emission integral of 10% CMA1 in TSPO1 with 1-(1/e) labelled as the characteristic luminescence lifetime. "Initial data taken at 300 K before cooling the film to 10 K, "Final" data upon warming back to 300 K after low-temperature measurements. Two activation energies are labelled, $E_Ak$ is extracted from the PL decay rate against temperature, and $E_APL$ is extracted from the integrated PL intensity against temperature. **e**, **f** Cryogenic time resolved PL peak position of 10% and 80% CMA1 in TSPO1 at 10 K, 150 K, and 300 K. Lines are results of Monte-Carlo simulation.

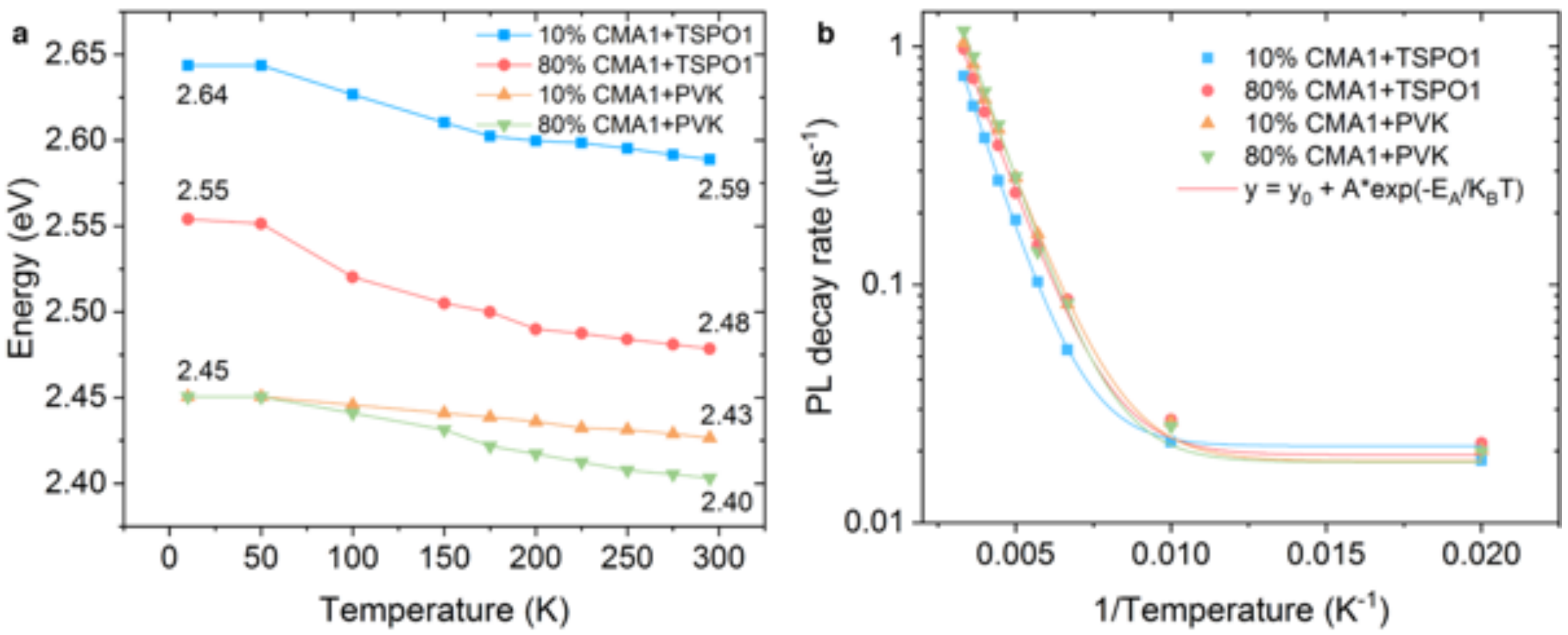

**Figure 5. a** Steady state PL peak energy of 10% and 80% concentration of CMA1 in TSPO1 and PVK host at different temperatures. PL blue shifts when decreasing the temperature. **b** PL decay rate of 10% and 80% concentration of CMA1 in TSPO1 and PVK at different temperatures as a function of 1/Temperature. PL decay rate is the reciprocal of characteristic luminescence lifetime from cryogenic emission integral. The fitted curves yield activation energies: $E_Ak$ (10% CMA1 + TSPO1) = 79 meV, $E_Ak$ (80% CMA1 + TSPO1) = 77 meV, $E_Ak$ (10% CMA1 + PVK) = 72 meV, $E_Ak$ (80% CMA1 + PVK) = 76 meV.

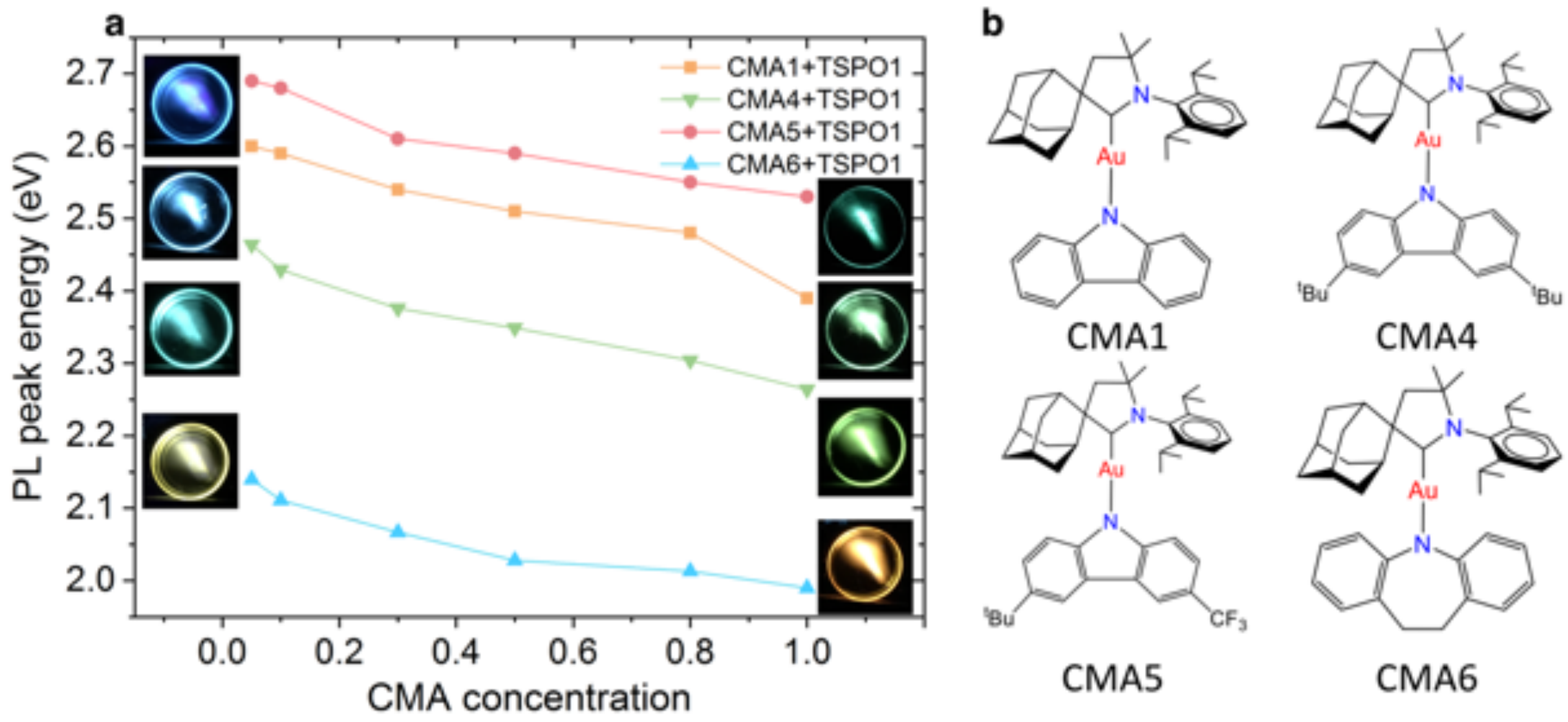

**Figure 6. a** Dependence of PL peak energy of various CMAs on the doping concentration in TSPO1 host. Photographs show photoluminescence of neat CMA1, CMA4, CMA5, and CMA6 and 5% weight concentration of CMAs in TSPO1 thin films under UV illumination. **b** Chemical structures of CMA1, CMA4, CMA5 and CMA6.

Supporting Information

**Environmental Control of Triplet Emission in Donor-Bridge-Acceptor Organometallics**

*Jiale Feng, Lupeng Yang, Alexander S. Romanov, Jirawit Ratanapreechachai, Saul T. E. Jones, Antti-Pekka M. Reponen, Mikko Linnolahti, Timothy J. H. Hele, Anna Köhler, Heinz Bässler, Manfred Bochmann, Dan Credgington**

# Supplementary Figures

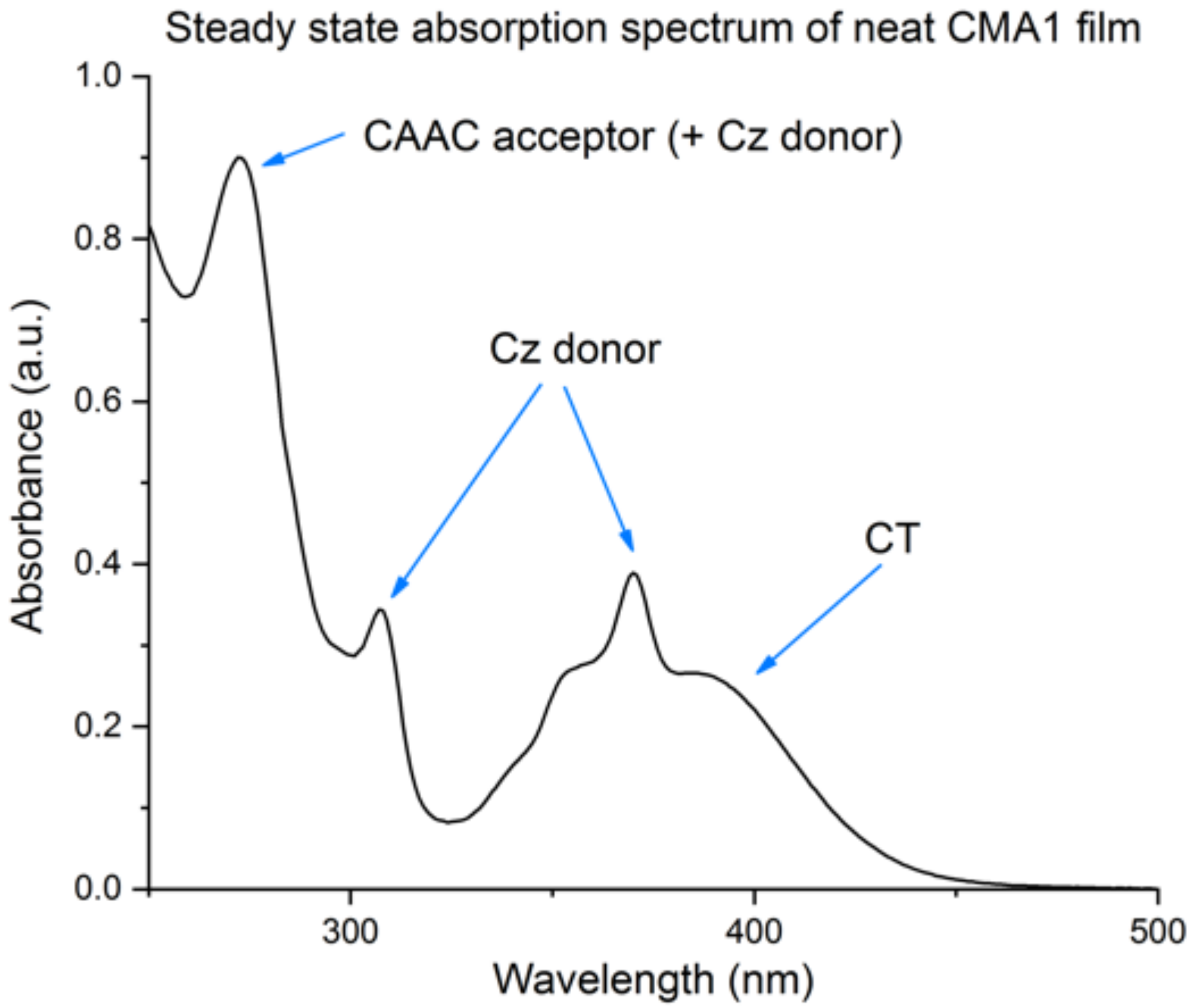

**Figure S1: Optical steady state absorption spectrum of CMA1 pristine thin film.** The film was spun from chlorobenzene solution (20 mg/ml) in a nitrogen glovebox. Vibronic progressions related to ligand-centred excitations of carbazole donor (Cz), CAAC acceptor, and the unstructured direct absorption to the $S_1$ CT state are labelled.

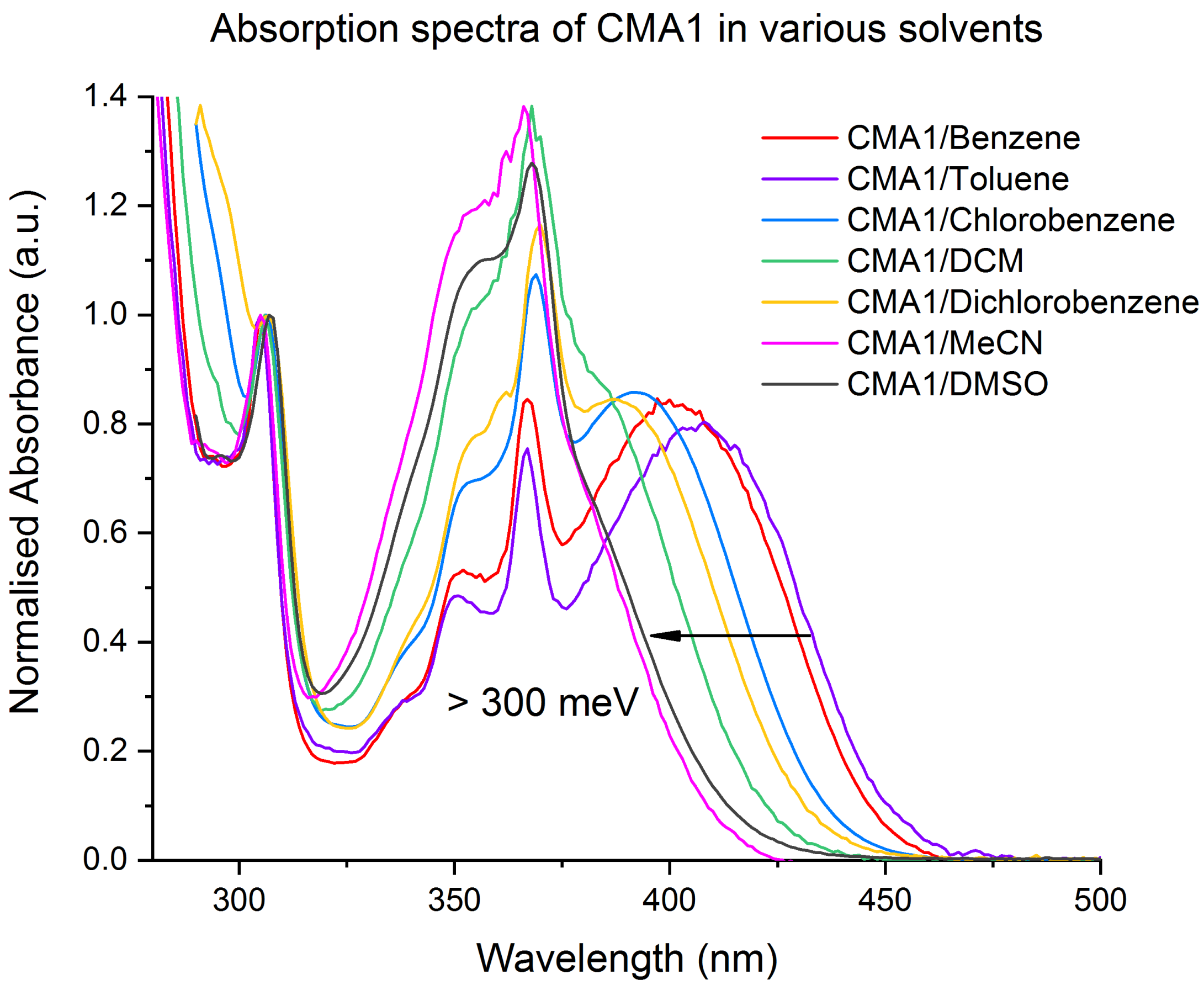

**Figure S2: Steady state absorption spectra of CMA1 in various solvents at 300 K.** Concentration is 1 mg/ml for all solutions. Spectra were normalised with respect to the carbazole absorption peak at around 305 nm. CT absorption peak blue shifts when increasing the solvent polarity, by over 300 meV for this solvent range.

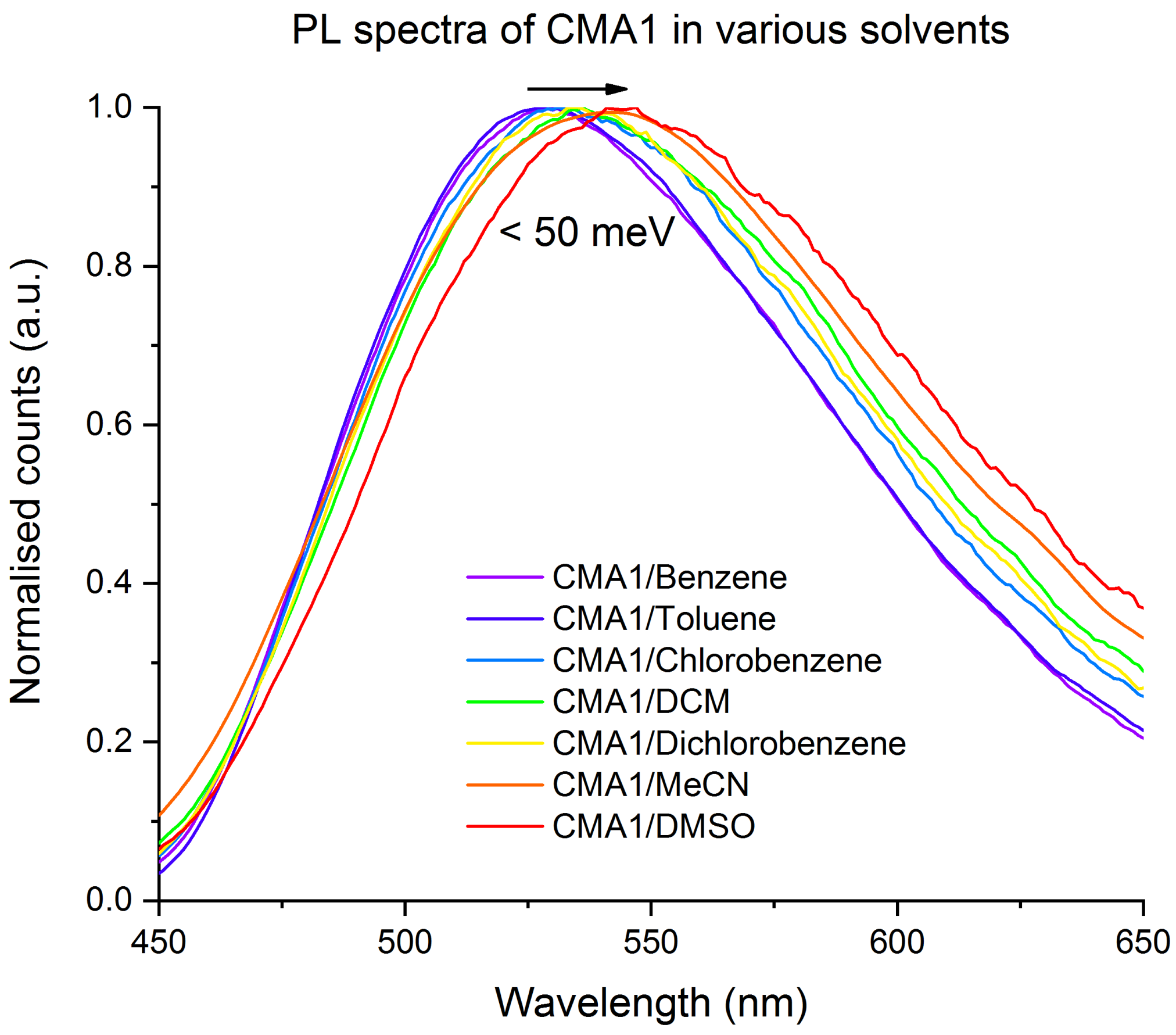

**Figure S3: Steady state photoluminescence spectra of CMA1 in various solvents at 300 K.** Concentration is 1 mg/ml for all solutions. Spectra were normalised with respect to their maximum. Emission peak position is weakly affected by solvent polarity, with shift smaller than 50 meV over this solvent range.

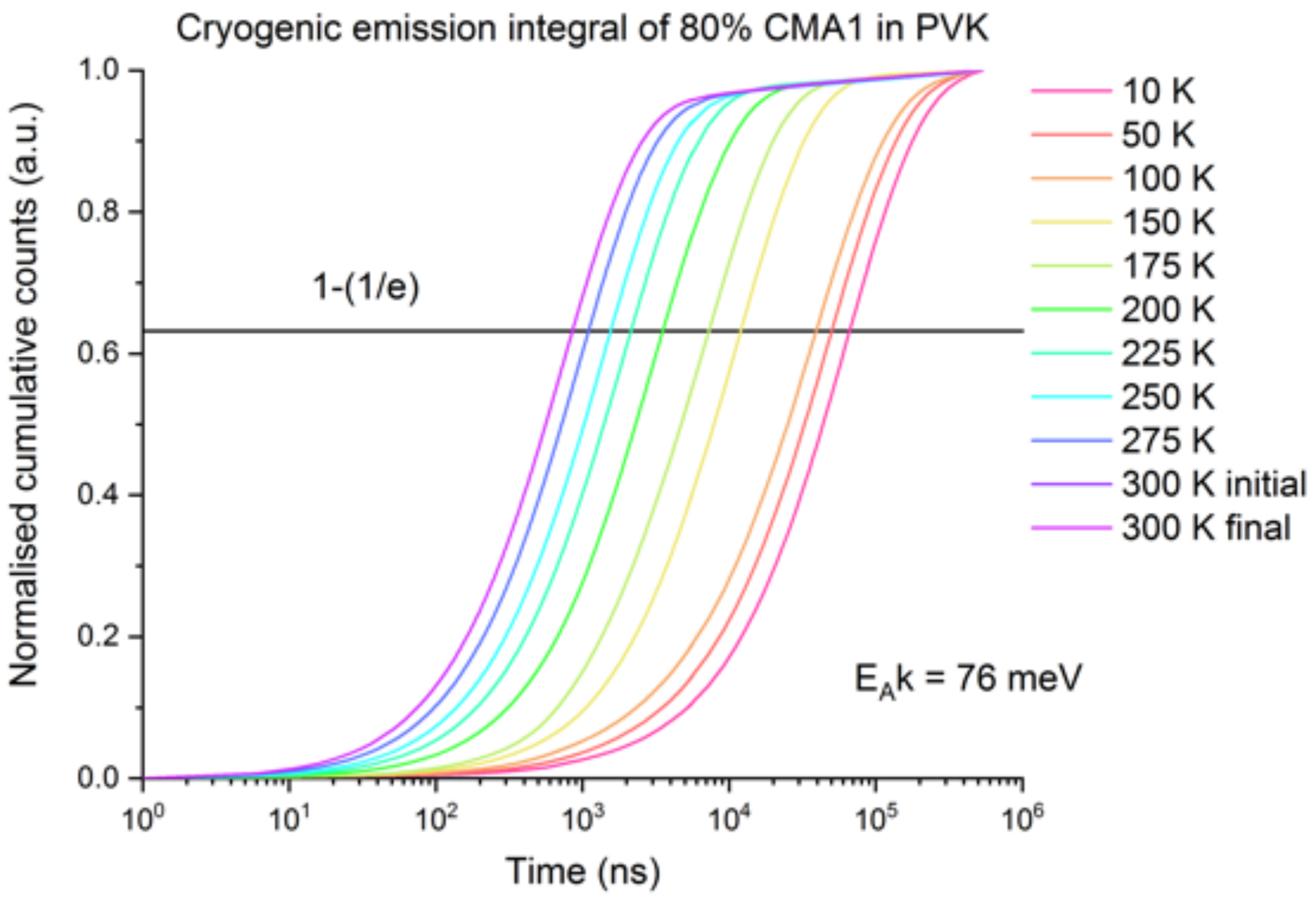

**Figure S4: Cryogenic emission integral of 80% CMA1 in PVK at different temperatures.** The film was spun from chlorobenzene solution (20 mg/ml) in a nitrogen glovebox and measured in helium environment. "Initial" data taken at 300 K before cooling the film to 10 K, "Final" data upon warming back to 300 K after low-temperature measurements. Characteristic activation energy extracted from PL decay rate is 76 meV.

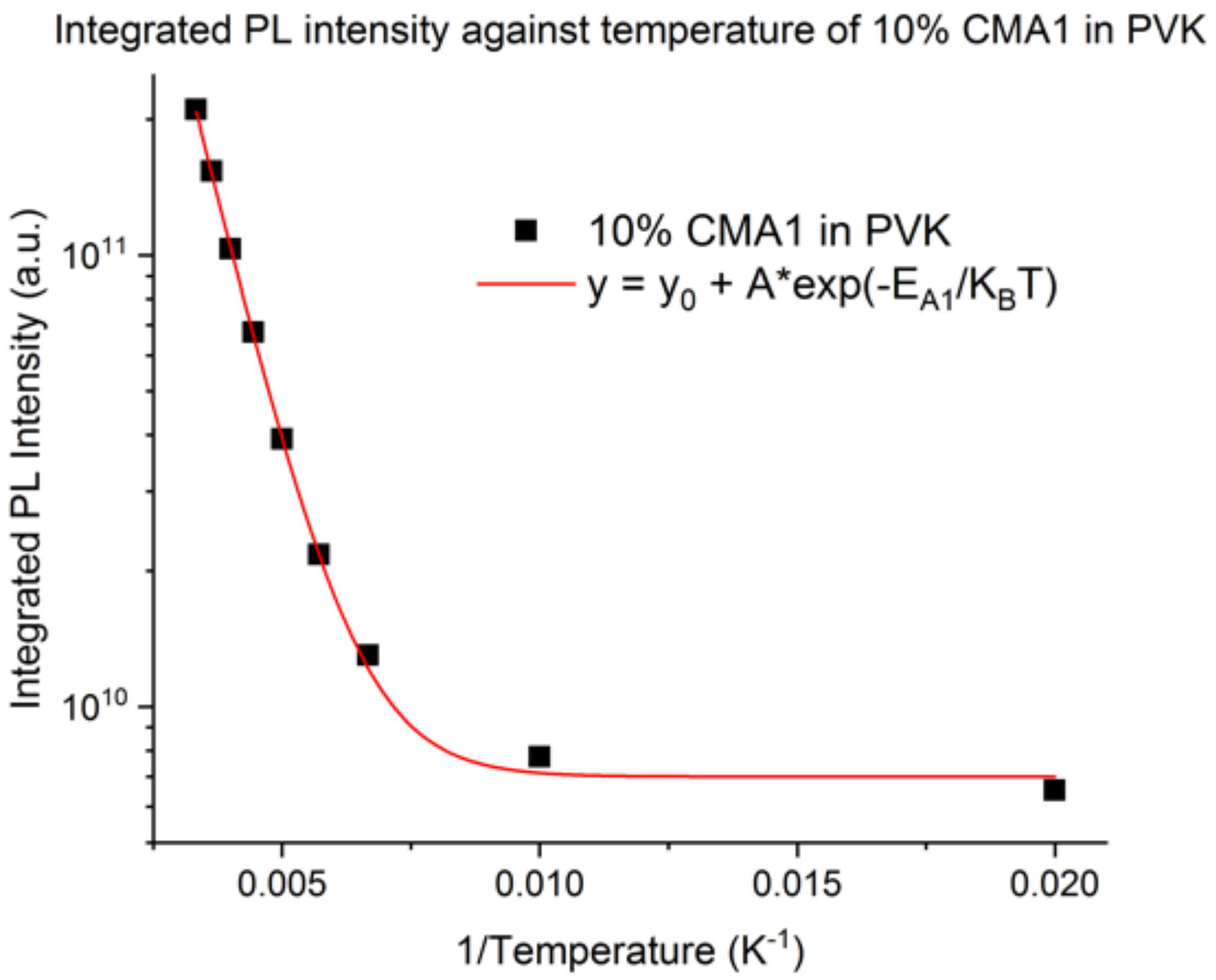

**Figure S5: Integrated PL intensity plotted against temperature of 10% CMA1 in PVK.** The integrated PL intensity is thermally activated and can be fitted by the equation shown. The activation energy extracted from the integrated PL intensity $\boldsymbol{E_A PL}$ = 70 meV.

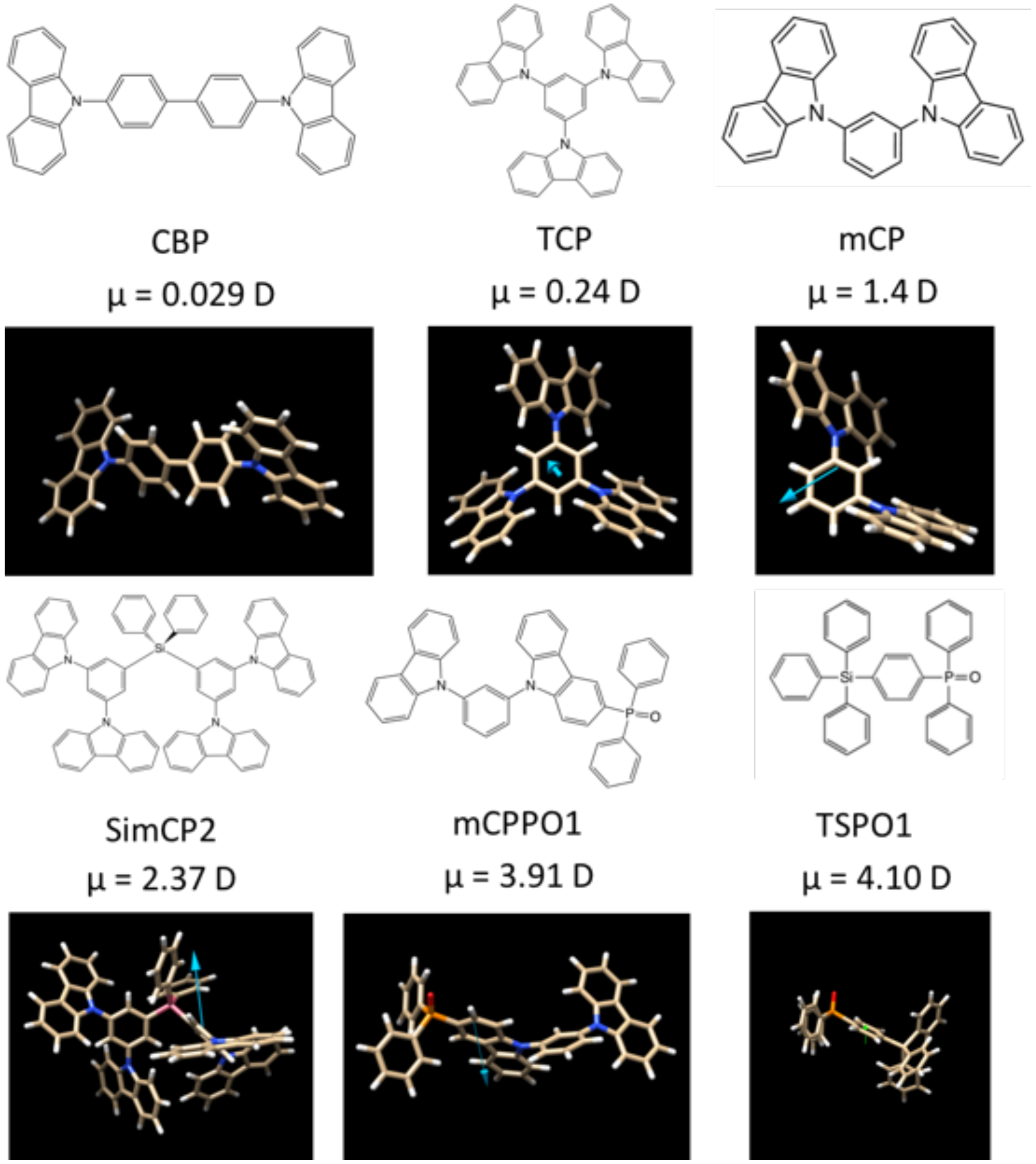

**Figure S6: Chemical structures and electric dipole moments of various small molecule hosts.** 4,4′–bis(N-carbazolyl)biphenyl (CBP); 1,3,5-Tris(*N*-carbazolyl)benzene (TCP); 1,3-Bis(N-carbazolyl)benzene (mCP); Bis[3,5-di(9H-carbazol-9-yl)phenyl]diphenylsilane (SimCP2); 9-(3-(9H-Carbazol-9-yl)phenyl)-3-(diphenylphosphoryl)-9H-carbazole (mCPPO1); Diphenyl-4-triphenylsilylphenyl-phosphine oxide (TSPO1). Molecular geometries and static dipole moments were calculated by DFT (B3LYP/6-31G**) with arrows indicating orientation of calculated dipole moments.

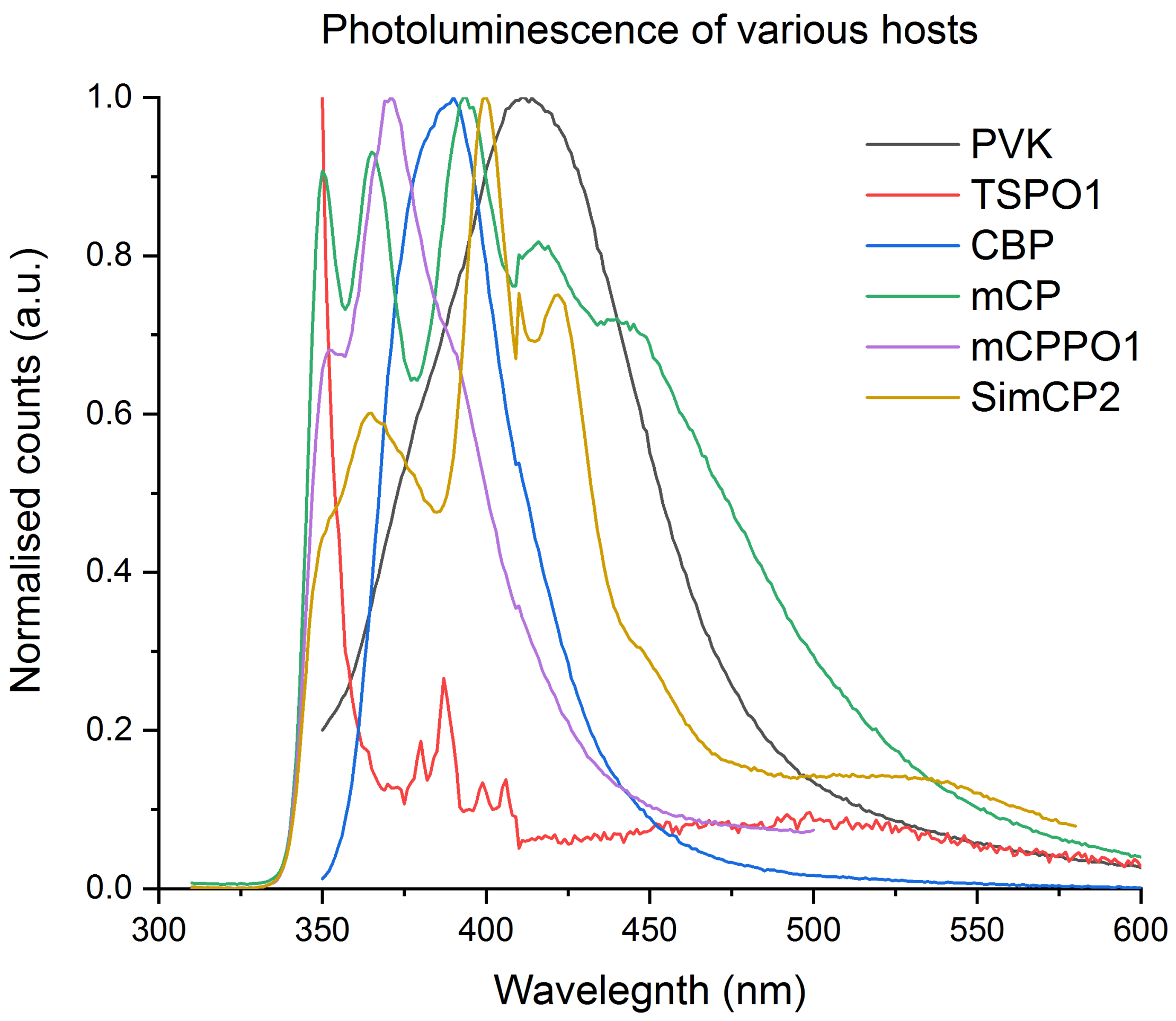

**Figure S7: Steady state photoluminescence of various hosts in solid thin films.** The films were spun from chlorobenzene solutions (20 mg/ml) in a nitrogen glovebox and measured by photoluminescence spectrometer, excited at 325 nm. The PL spectra of neat host molecules show that the blue shifted PL in host-guest composite films are from the guest emission, not from the host.

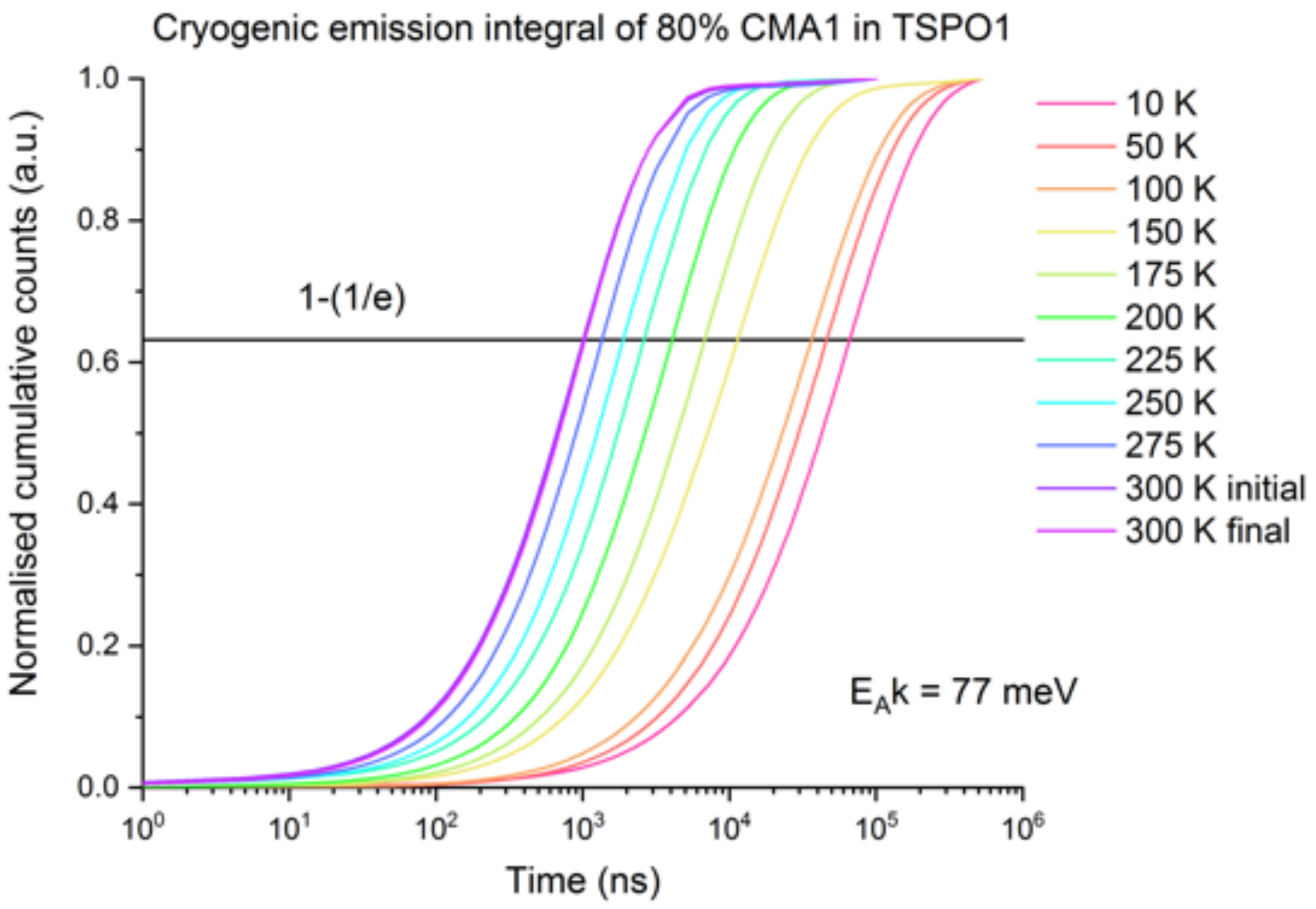

**Figure S8: Cryogenic total emission integral of 80% CMA1 in TSPO1 at different temperatures.** The film was spun from chlorobenzene solution (20 mg/ml) in a nitrogen glovebox and measured in helium environment. “Initial” data taken at 300 K before cooling the film to 10 K, “Final” data upon warming back to 300 K after low-temperature measurements.

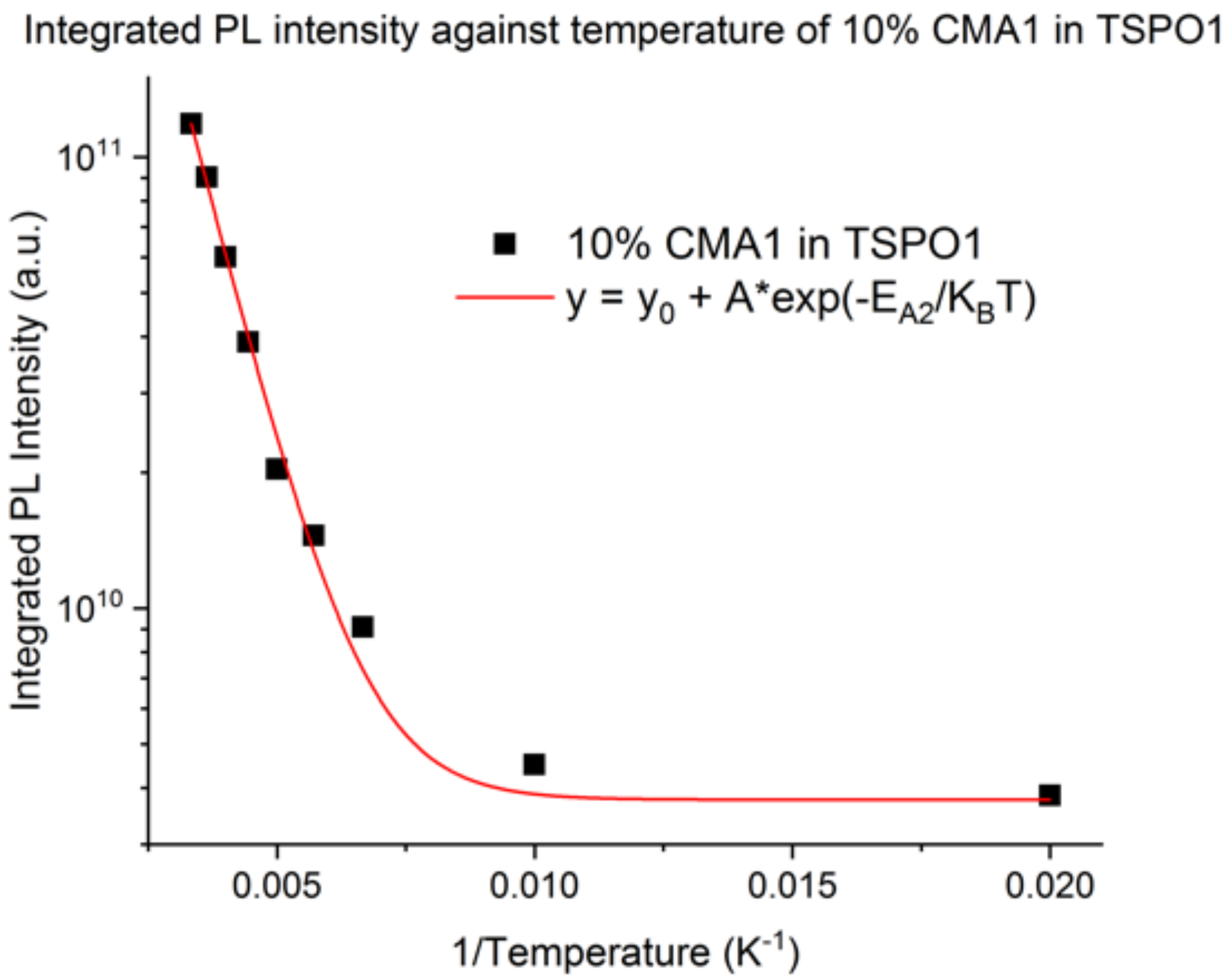

**Figure S9: Integrated PL intensity plotted against temperature of 10% CMA1 in TSPO1.** The integrated PL intensity is thermally activated and can be fitted by the equation shown. The activation energy extracted from the integrated PL intensity $\boldsymbol{E_A PL}$ = 73 meV.

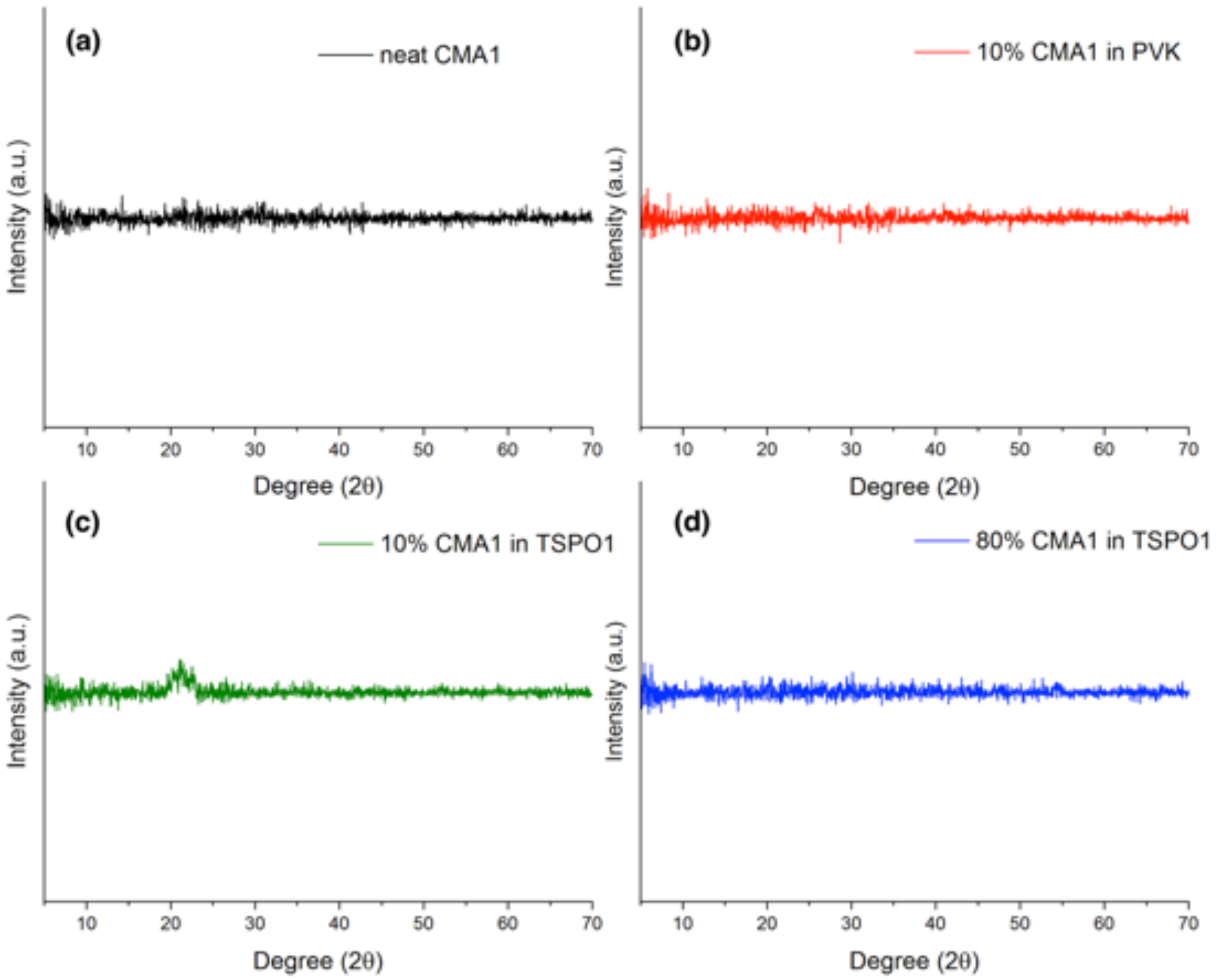

**Figure S10: XRD patterns of: (a) neat CMA1; (b) 10 wt.% CMA1 in PVK; (c) 10 wt.% CMA1 in TSPO1; (d) 80 wt.% CMA1 in TSPO1 thin films.** The intensity of X-ray scatter is presented on the same scale. The films were spun from chlorobenzene solutions (20 mg/ml) in a nitrogen glovebox. The feature around 20 degrees in panel (c) is from TSPO1 host.

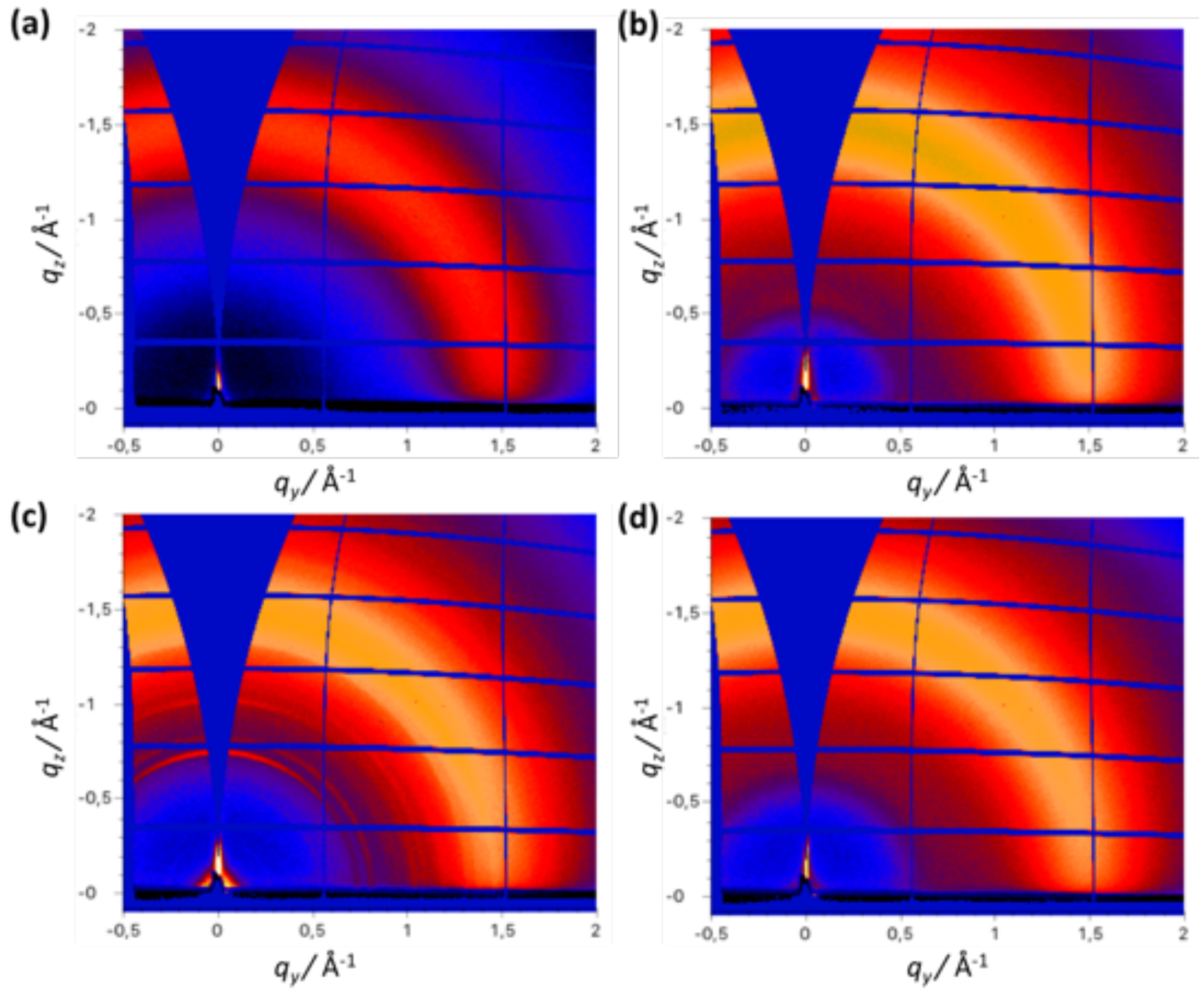

**Figure S11: 2D GIWAXS patterns of: (a) neat CMA1; (b) 10 wt.% CMA1 in PVK; (c) 10 wt.% CMA1 in TSPO1; (d) 80 wt.% CMA1 in TSPO1 thin films.** The intensity of X-ray scatter is presented on a common logarithmic colour scale. The films were spun from chlorobenzene solutions (20 mg/ml) in a nitrogen glovebox. The features seen in panel (c) are from TSPO1 host.

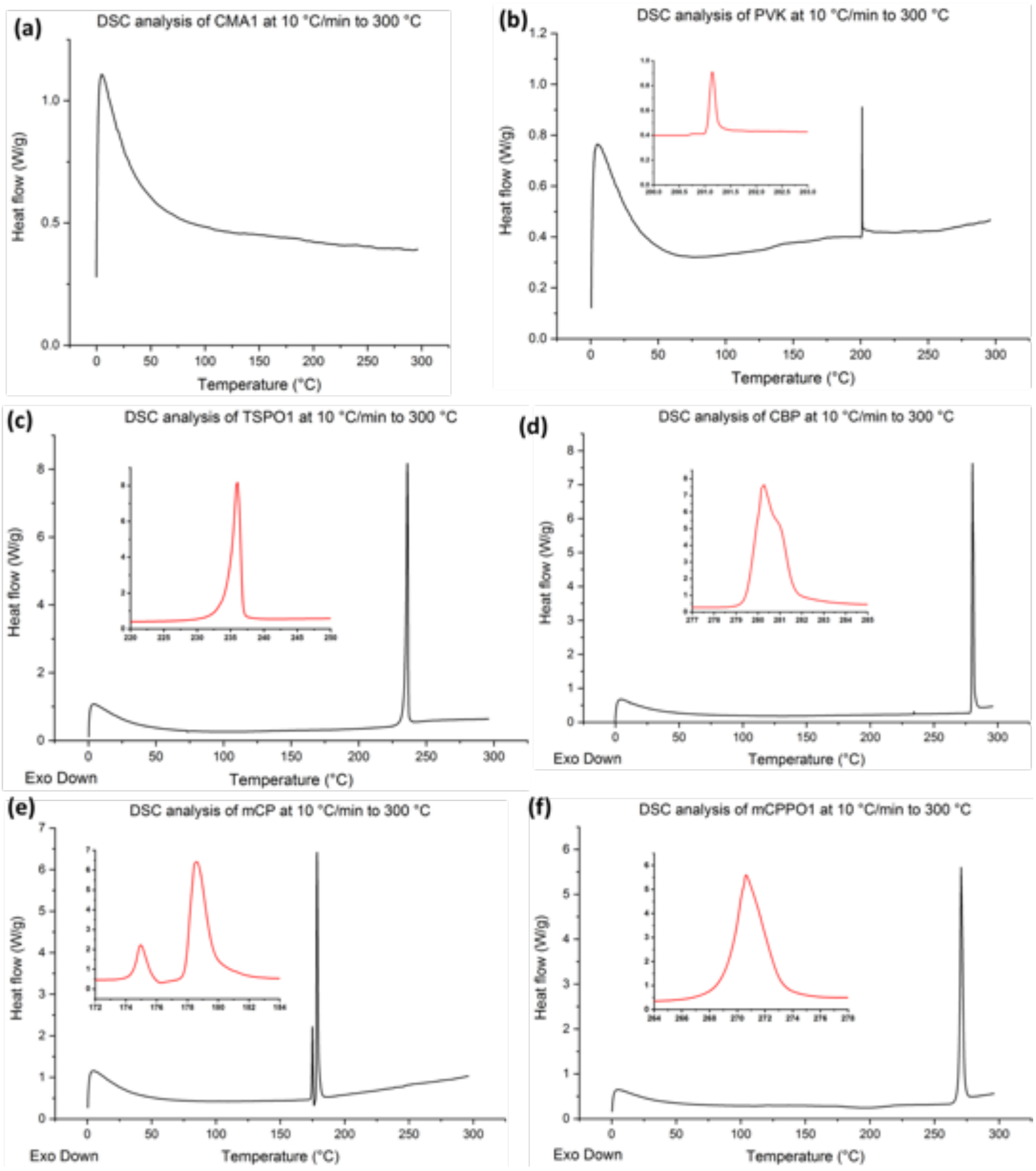

**Figure S12: DSC analysis of: (a) CMA1; (b) PVK; (c) TSPO1; (d) CBP; (e) mCP; (f) mCPPO1 neat powder.** Heat rate is 10 ˚C/min. Samples were measured under nitrogen environment.

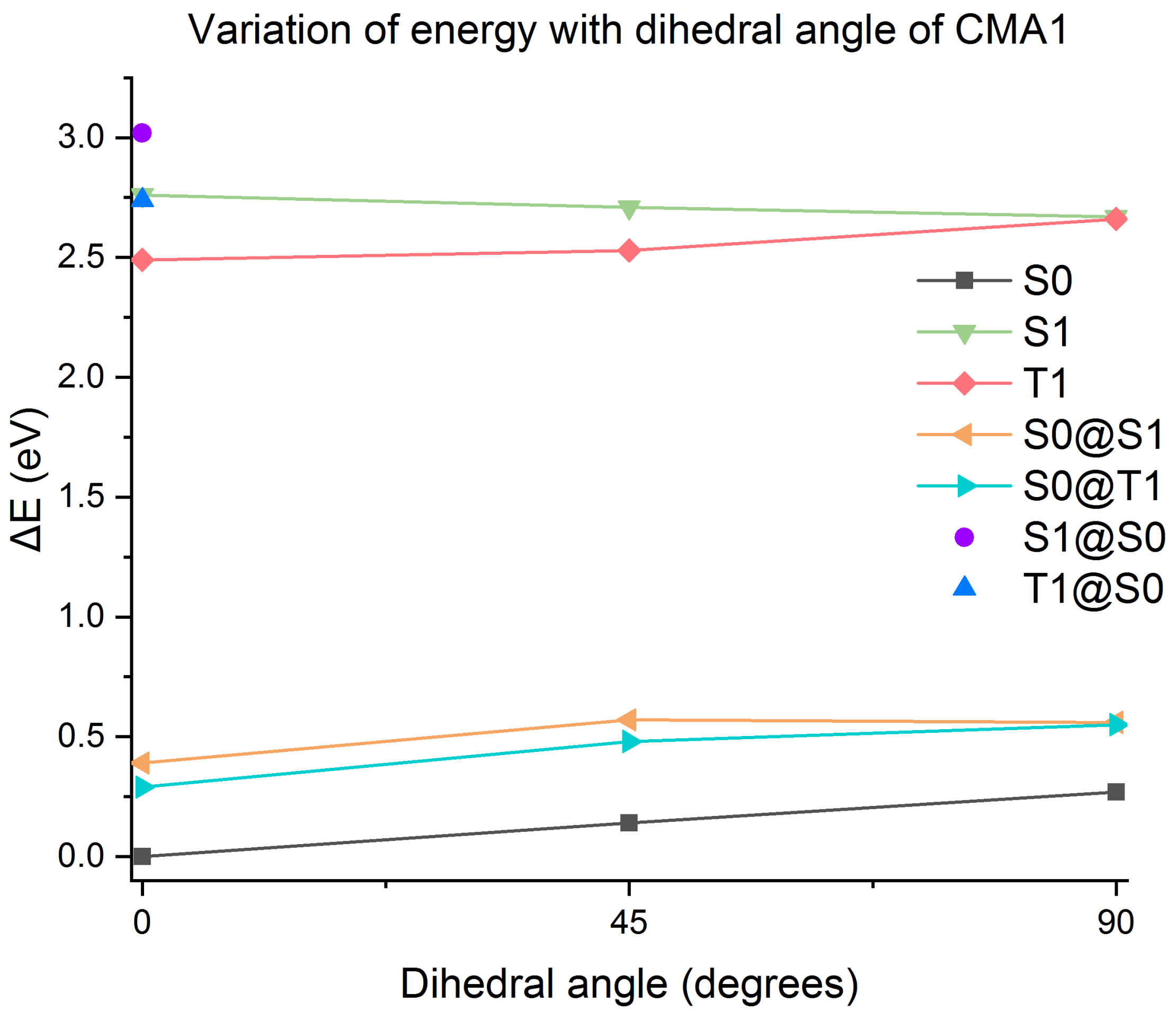

**Figure S13: Energy diagram for CMA1 with various orientation between the planes of CAAC carbene and carbazole: planar (0º), twisted (45º) and rotated (90º).** S0 stands for ground state geometry; S1 and T1 are relaxed geometries for S1 and T1 excited states; S0@S1 stands for S0 ground state with geometry of S1 excited state; S0@T1 stands for S0 ground state with geometry of T1 excited state; S1@S0 stands for S1 excited state with geometry of S0 ground state; T1@S0 stands for T1 excited state with geometry of S0 ground state. All S0 were calculated with DFT and excited states were calculated with time-dependent DFT (TD-DFT) using the MN15/def2-TZVP method. The energies were summarised in the Table S8.

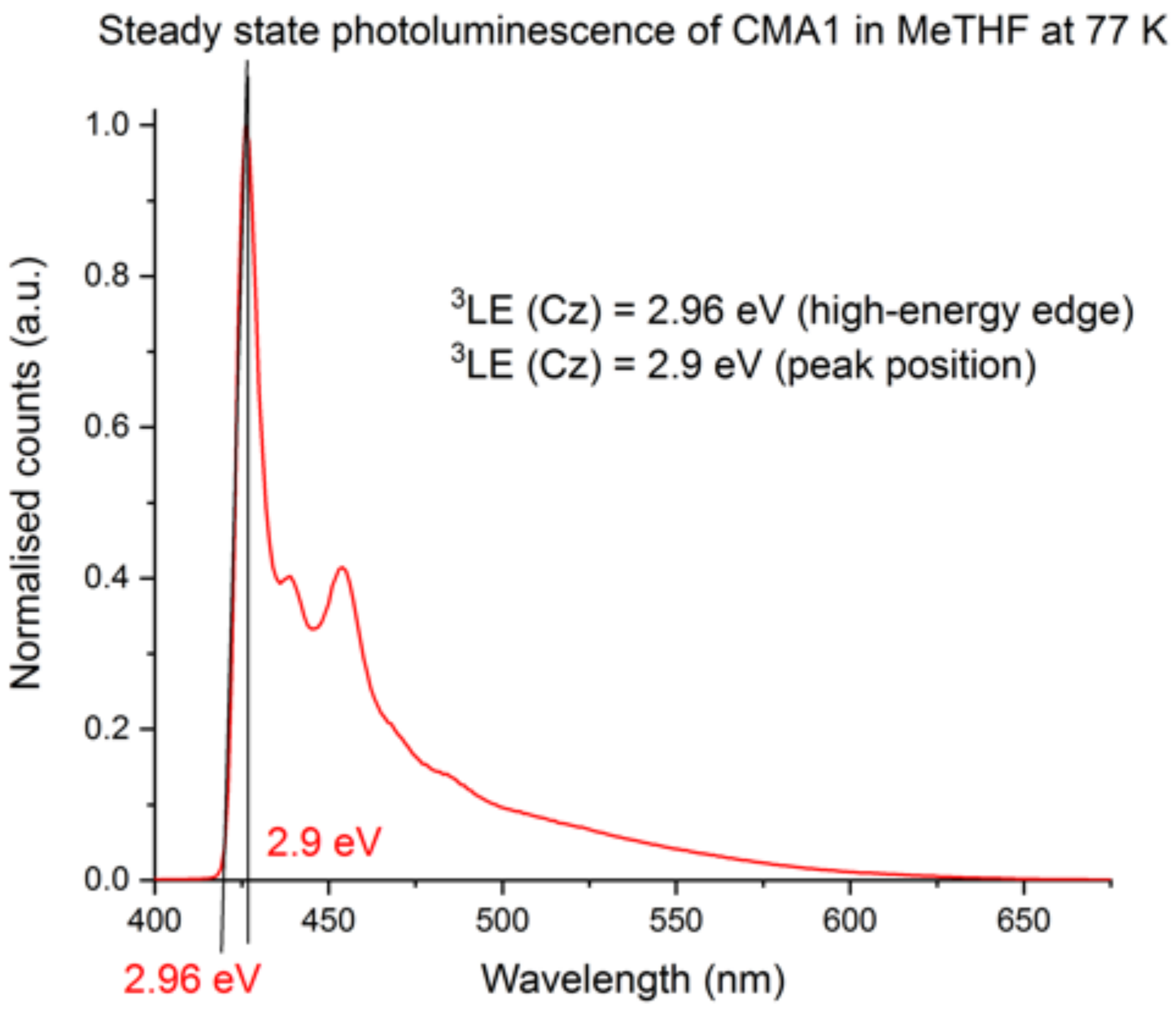

**Figure S14: Steady state photoluminescence of CMA1 in MeTHF solvent at 77 K.** Structured PL is from the triplet localised to the carbazole donor (Cz). Solution was deoxygenised and sealed in cuvette. Concentration is 1 mg/ml.

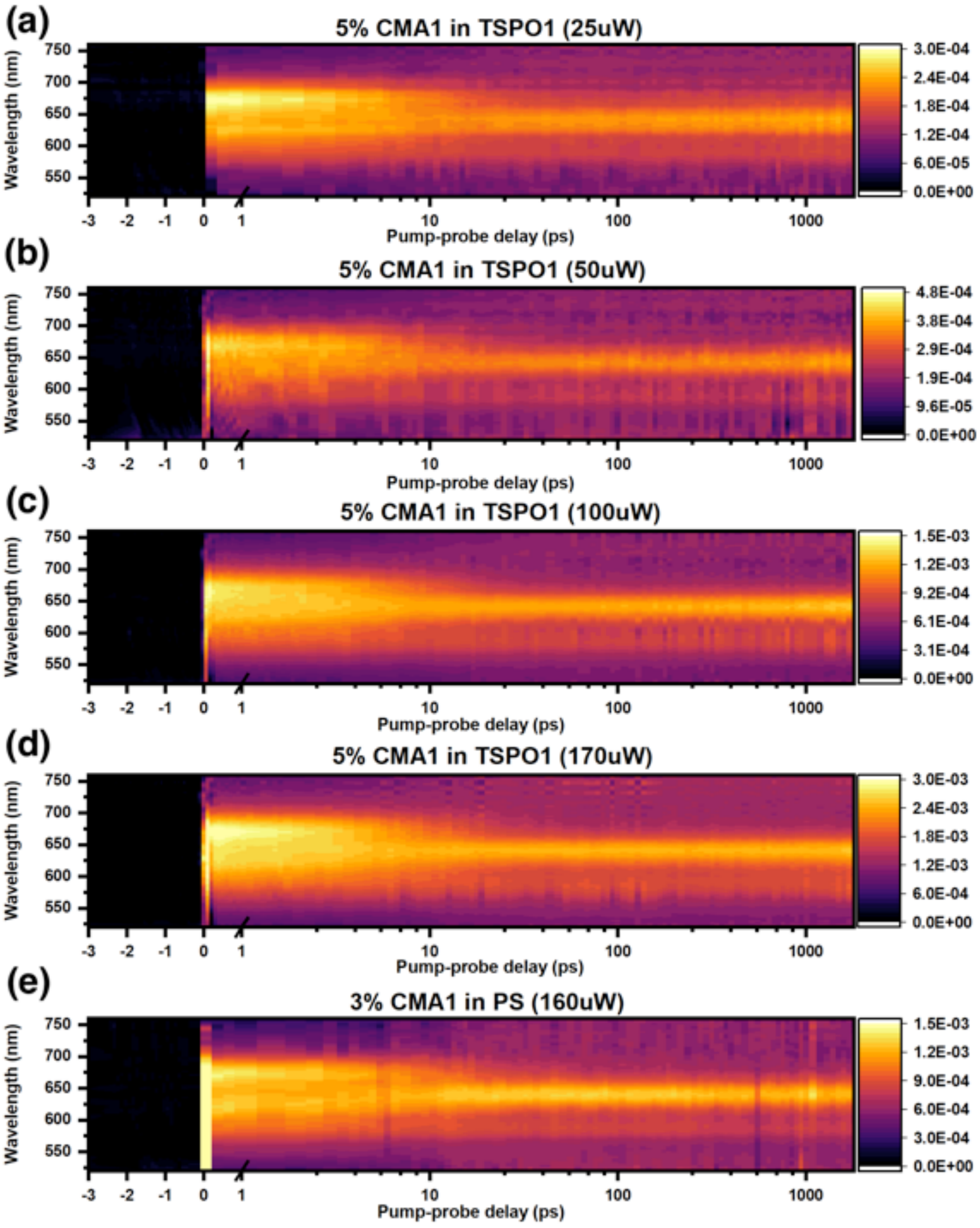

**Figure S15: (a) - (d) Transient absorption (TA) colour map of 5 wt.% CMA1 in TSPO1 films under various pump power, 25 µW, 50 µW, 100 µW, and 170 µW; (e) 3 wt.% CMA1 in polystyrene film under 160 µW pump power.** The initial excited state absorption associated with singlet is located at 680 nm, which fades away at early times around 5 to 6 ps. The narrower peak located at 645 nm is associated with triplet, which exists at longer timescale.

The intersystem crossing (ISC) time of each sample is estimated by the crossover of singlet and triplet kinetics, and is summarised in **Table S9**.

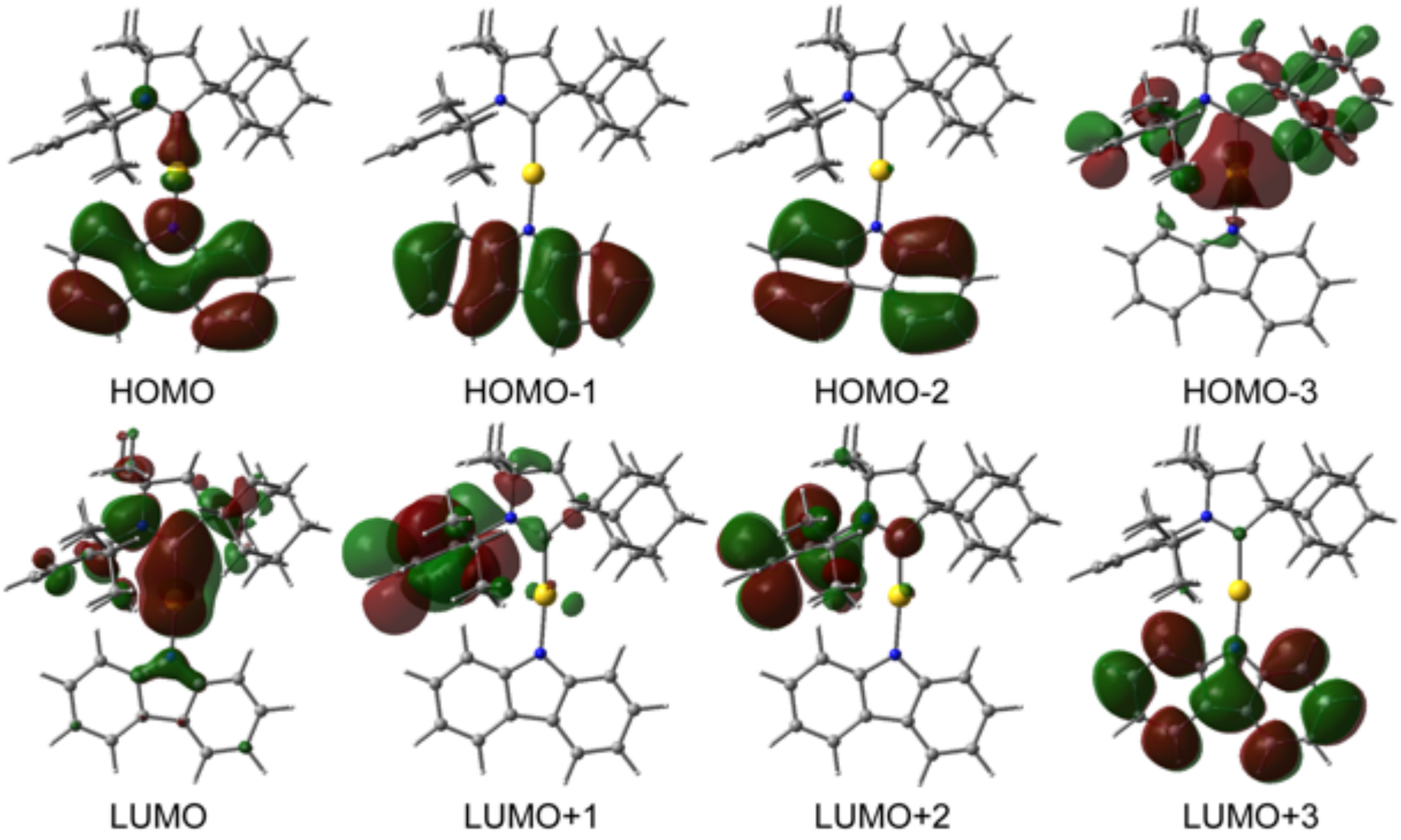

**Figure S16: Molecular orbitals of CMA1 from DFT calculations, red/green corresponds to positive/negative sign of wavefunctions.**

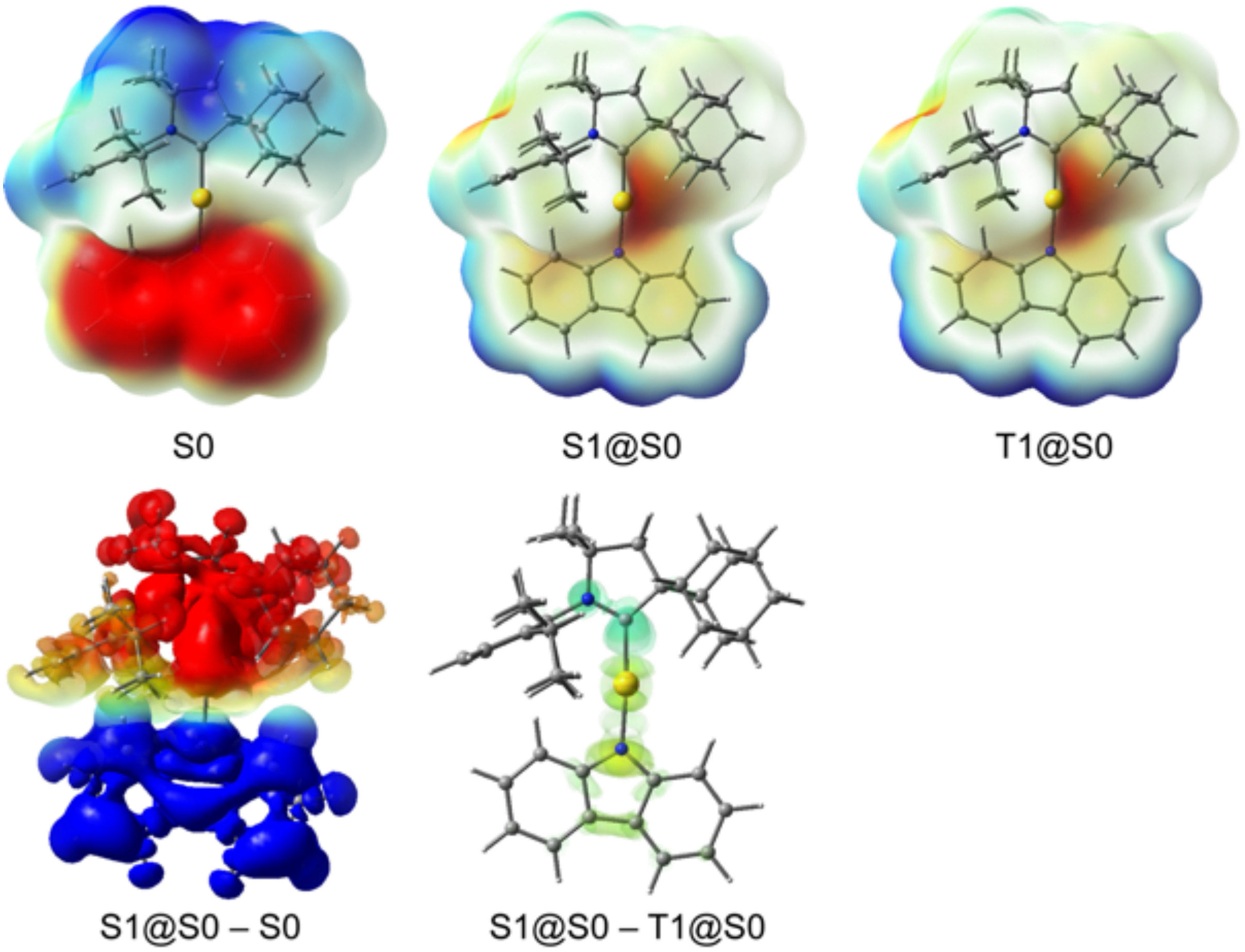

**Figure S17: Electrostatic potential maps of CMA1 at different states.** S0 stands for ground state geometry; S1@S0 stands for S1 excited state with geometry of S0 ground state; T1@S0 stands for T1 excited state with geometry of S0 ground state.

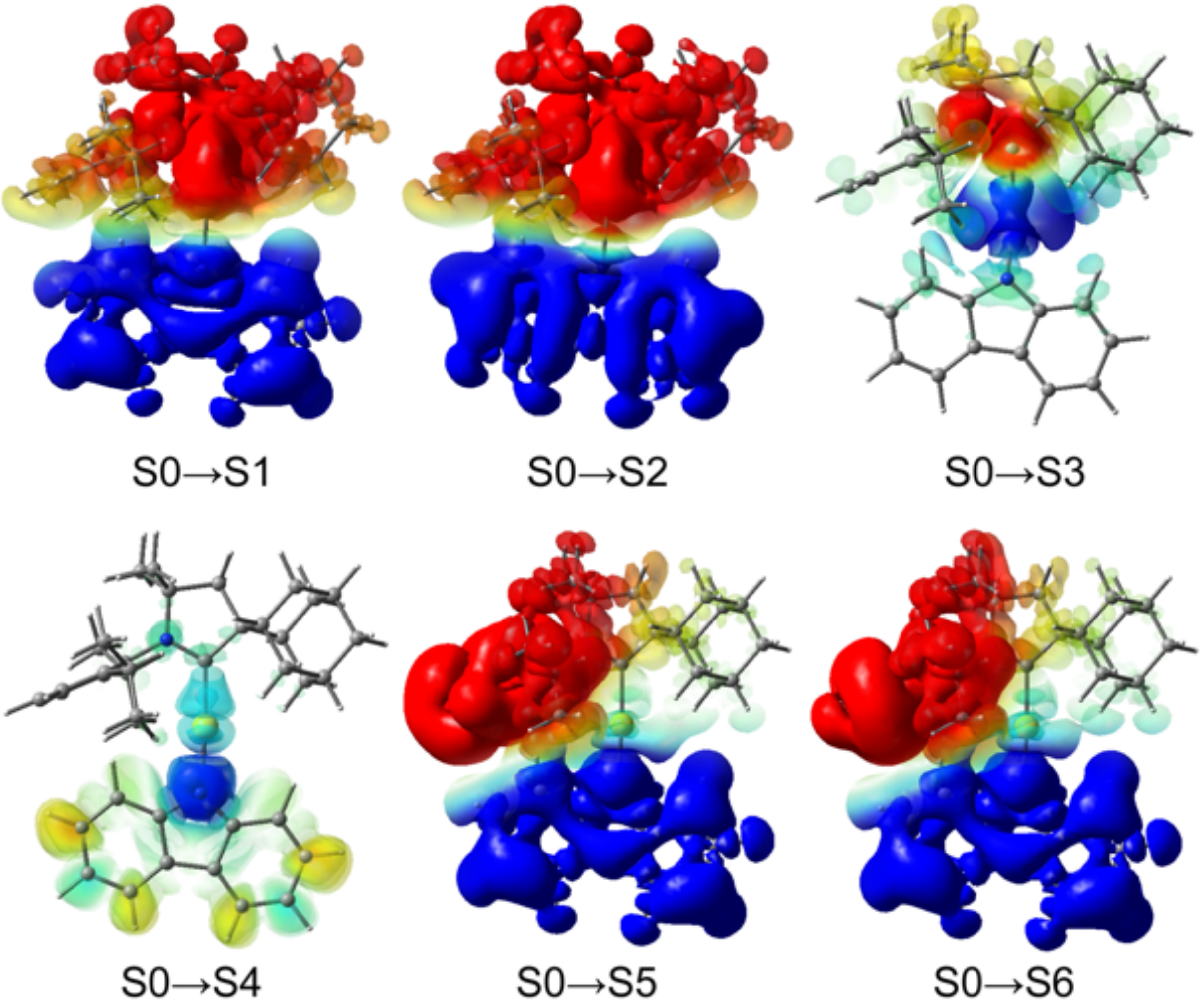

**Figure S18: Charge transfer upon excitation, from blue region to red region. Orbital contributions to vertical excitations are summarised in Table S10.**

# Supplementary Tables

| CMA1 / PVK | Luminescence Lifetime | PLQE |
|---|---|---|
| **5% CMA1** | 1040 ns | 67% |
| **10% CMA1** | 1000 ns | 75% |
| **30% CMA1** | 920 ns | 76% |
| **50% CMA1** | 850 ns | 94% |
| **80% CMA1** | 840 ns | 81% |
| **100% CMA1** | 970 ns | 80% |

**Table S1: Luminescence lifetime and Photoluminescence quantum efficiency (PLQE) of CMA1 in PVK host varying concentrations from 5 wt.% to neat CMA1 films.** Luminescence lifetime was measured in vacuum with sample excited by 400 nm laser. Luminescence lifetime was determined by the time when the emission reaches 1-(1/e) of the total time-integrated emission, see Figure 2b. PLQE was measured in a nitrogen environment in an integrating sphere with sample excited by 405 nm laser of 1 mW power. PLQE was calculated based on the De Mello method.

| CMA1 / TSPO1 | Luminescence Lifetime | PLQE |
|---|---|---|
| **5% CMA1** | 1400 ns | 65% |
| **10% CMA1** | 1330 ns | 72% |
| **30% CMA1** | 1190 ns | 78% |
| **50% CMA1** | 1180 ns | 70% |
| **80% CMA1** | 1010 ns | 75% |
| **100% CMA1** | 970 ns | 80% |

**Table S2: Luminescence lifetime and Photoluminescence quantum efficiency (PLQE) of CMA1 in TSPO1 host varying concentrations from 5 wt.% to neat CMA1 films.** Luminescence lifetime was measured in vacuum with sample excited by 400 nm laser. Luminescence lifetime was determined by the time when the emission reaches 1-(1/e) of the total time-integrated emission, see Figure 4b. PLQE was measured in a nitrogen environment in an integrating sphere with sample excited by 405 nm laser of 1 mW power. PLQE was calculated based on the De Mello method.

| CMA1 / Solvents | Luminescence Lifetime | PLQE |
|---|---|---|
| **CMA1/tol** | 1250 ns | 98% |
| **CMA1/CB** | 1100 ns | 98% |
| **CMA1/DCB** | 1000 ns | 95% |
| **CMA1/DMSO** | 1050 ns | 77% |

**Table S3: Luminescence lifetime and Photoluminescence quantum efficiency (PLQE) of CMA1 in different solvents (1 mg/ml) deoxygenised and sealed in cuvettes.** Luminescence lifetime was measured under 400 nm laser. Luminescence lifetime was determined by the time when the emission reaches 1-(1/e) of the total time-integrated emission. PLQE was measured in an integrating sphere with sample excited by 405 nm laser of 1 mW power. PLQE was calculated based on the De Mello method.

| CMA1 / PVK (300 K) | Mean energy (eV) | Deviation $\sigma$ (eV) |
|---|---|---|
| **5% CMA1** | 2.427 | 0.048 |
| **10% CMA1** | 2.427 | 0.048 |
| **30% CMA1** | 2.427 | 0.048 |
| **50% CMA1** | 2.427 | 0.048 |
| **80% CMA1** | 2.427 | 0.048 |
| **100% CMA1** | 2.427 | 0.048 |

**Table S4: Mean energy and deviation of density of states (DOS) input in Monte-Carlo simulations of CMA1 in PVK concentration series at 300 K.** These parameters correspond to the simulations in Figure 2c, with triplet states concentration being the only variable. The spectral diffusion becomes slower upon the dilution of CMA1, however, the relaxation stays relatively constant as the concentration changes from 100% to 5%. Note that as the concentration approaches 0, the simulation result quickly becomes a straight line.

| CMA1 / TSPO1 (300 K) | Mean energy (eV) | Deviation $\sigma$ (eV) |
|---|---|---|
| **5% CMA1** | 2.54 | 0.023 |
| **10% CMA1** | 2.528 | 0.028 |
| **30% CMA1** | 2.515 | 0.036 |
| **50% CMA1** | 2.504 | 0.042 |
| **80% CMA1** | 2.49 | 0.044 |
| **100% CMA1** | 2.427 | 0.048 |

**Table S5: Mean energy and deviation of density of states (DOS) input in Monte-Carlo simulations of CMA1 in TSPO1 concentration series at 300 K.** In contrast to the table above, both the Gaussian mean energy and width vary for different concentrations. These parameters correspond to the simulations in Figure 4c**.**

| CMA1 / PVK (cryo) | Mean energy (eV) | Deviation $\sigma$ (eV) |
|---|---|---|
| **10% CMA1 (10 K)** | 2.49 | 0.048 |
| **10% CMA1 (150 K)** | 2.445 | 0.048 |
| **80% CMA1 (10 K)** | 2.485 | 0.048 |
| **80% CMA1 (150 K)** | 2.45 | 0.048 |

**Table S6: Mean energy and deviation of density of states (DOS) input in Monte-Carlo simulations of 10% and 80% CMA1 in PVK at 10 K and 150 K.** These parameters correspond to the simulations in Figure 2e, f.

| CMA1 / TSPO1 (cryo) | Mean energy (eV) | Deviation $\sigma$ (eV) |
|---|---|---|
| **10% CMA1 (10 K)** | 2.62 | 0.028 |
| **10% CMA1 (150 K)** | 2.569 | 0.028 |
| **80% CMA1 (10 K)** | 2.558 | 0.044 |
| **80% CMA1 (150 K)** | 2.528 | 0.044 |

**Table S7: Mean energy and deviation of density of states (DOS) input in Monte-Carlo simulations of 10% and 80% CMA1 in TSPO1 at 10 K and 150 K.** These parameters correspond to the simulations in Figure 4e, f. As temperature reaches 0 K, the excitation becomes trapped and the simulation energy approaches a constant.

| $\Delta E$ (eV) | S0 | S1@S0 | T1@S0 | S1 | Oscillation strength | T1 | S0@S1 | S0@T1 | Fluorescence= S1-S0@S1 | Phosphorescence= T1-S0@T1 |
|---|---|---|---|---|---|---|---|---|---|---|
| **Planar** | 0.00 | 3.02 | 2.74 | 2.76 | 1.00 | 2.49 | 0.39 | 0.29 | 2.37 = 523 nm | 2.20 = 563 nm |
| **Twisted (45°)** | 0.14 | | | 2.71 | 0.46 | 2.53 | 0.57 | 0.48 | 2.14 = 580 nm | 2.05 = 604 nm |
| **Rotated** | 0.27 | | | 2.67 | 0.00 | 2.66 | 0.56 | 0.55 | 2.11 = 586 nm | 2.11 = 589 nm |

**Table S8：The table summarises the energies of CMA1 with various orientations in Figure S13.** Oscillation strength is calculated relative to planar orientation.

| | CMA1:TSPO1 5wt% 25 µW | CMA1:TSPO1 5wt% 50 µW | CMA1:TSPO1 5wt% 100 µW | CMA1:TSPO1 5wt% 170 µW | CMA1:PS 3wt% 160 µW |
|---|---|---|---|---|---|
| **Estimated Intersystem crossing (ISC) time (ps)** | 5.6 | 5.6 | 5.6 | 5.3 | 5.5 |

**Table S9: Estimated intersystem crossing (ISC) time extracted from Figure S15.**

| State | Energy (eV) | Main contributions | Oscillator strength | HONTO | LUNTO |
|---|---|---|---|---|---|
| **S1** | 3.02 | H → L 98% | 0.1556 | ≈ HOMO | ≈ LUMO |
| **S2** | 3.68 | H-1 → L 99% | 0.0004 | ≈ HOMO-1 | ≈ LUMO |
| **S3** | 3.93 | H-3 → L 91%<br>H-6 → L 2% | 0.0033 | ≈ HOMO-3 | ≈ LUMO |
| **S4** | 3.94 | H → L+3 91%<br>H-1 → L+7 4%<br>H → L+1 3% | 0.0490 | ≈ HOMO | ≈ LUMO+3 |
| **S5** | 4.03 | H → L+1 95%<br>H → L+3 3% | 0.0035 | ≈ HOMO | ≈ LUMO+1 |
| **S6** | 4.17 | H → L+2 98% | 0.0026 | ≈ HOMO | ≈ LUMO+2 |

**Table S10: Orbital contributions to vertical excitations (S0 → S1-S6).**

# Supplementary Text

## 1. Thermally activated decay rate

$$k_{decay} = k_r + k_{nr} = k_{r0} * \exp\left(-\frac{E_{A1}}{k_B T}\right) + k_{nr0} * \exp\left(-\frac{E_{A2}}{k_B T}\right) \approx k_{r0} * \exp\left(-\frac{E_{A1}}{k_B T}\right) \quad (S1)$$

$$PL_{total} = \int_0^\infty k_r dt = PL_0 * \exp\left(-\frac{E_{A3}}{k_B T}\right) \quad (S2)$$

$k_{decay}$: total decay rate of excitons

$k_r$: radiative decay rate of excitons

$k_{nr}$: non-radiative decay rate of excitons

$k_{r0}$: radiative decay rate constant

$k_{nr0}$: non-radiative decay rate constant

$E_{A1}$: activation energy of radiative decay

$E_{A2}$: activation energy of non-radiative decay

$PL_{total}$: total time-integrated emission

$PL_0$: total time-integrated emission constant

$E_{A3}$: activation energy of photoluminescence

$k_B$: Boltzmann constant

$T$: temperature in Kelvin

The total decay rate comprises radiative decay and non-radiative decay, both of which can be assumed to be thermally activated, with activation energy of $E_{A1}$ and $E_{A2}$. The total time-integrated emission represents the integral of the radiative decay over time, which is also thermally activated with activation energy of $E_{A3}$. Given that $E_{A1}$ is measured to be the same as $E_{A3}$, the thermal activation is primarily of the radiative triplet decay rate.

## 2. Non-adiabatic Marcus-type hopping rate between two sites

$$W_{ij} = \frac{J_0^2 \exp\left(-2\frac{r_{ij}}{L}\right)}{\hbar}\sqrt{\frac{\pi}{4E_a k_B T}}\exp\left[-\frac{(4E_a + \varepsilon_j - \varepsilon_i)^2}{16E_a k_B T}\right] \qquad (S3)$$

$W_{ij}$: hopping rate from site $i$ to site $j$

$J_0$: electronic coupling between triplet sites

$r_{ij}$: distance between site $i$ and site $j$

$L$: wavefunction overlap

$E_a$: activation energy

$\varepsilon_i$ and $\varepsilon_j$: energies of site $i$ and site $j$

$k_B$: Boltzmann constant

$T$: temperature in Kelvin

$\hbar$: reduced Planck constant

Hopping probability from site $i$ to site $j$

$$P_{ij} = \frac{W_{ij}}{\sum_{k\neq i} W_{ik}} \qquad (S4)$$

Hopping time

$$t = \frac{x}{\sum_{k\neq i} W_{ik}} \qquad (S5)$$

$x$: A random number drawn from an exponential distribution (PDF: $f(x) = e^{-x}$)

## Parameterisation of the Monte-Carlo model

The lattice constant in all Monte-Carlo simulations was set to 1 nm. In addition, the following parameters were used for Figure 2 and Figure 4: $J_0 = 1.2$ meV, $L = 0.2$ nm and $\lambda = 240$ meV. This parameter set gave an intrinsic jumping frequency ~50 $ns^{-1}$. A constant off-set of 87 meV between the Gaussian mean energy and the initial excitation was chosen so that the excitation population was allowed to vary as the shape of DOS changes. The fitting strategy was as follows. A Gaussian mean energy and deviation of 2.427 eV and 0.048 eV were selected to reproduce the pristine data in Figure 2c and Figure 4c. Random sites in the lattice were then blocked to fit other concentrations. The concentration of active sites did not match experimental concentration by weight as the exact density due to the possible inhomogeneous local microstructure. However, the concentration by volume values were close to the ones calculated by assuming guests and hosts having the same density in the films as of pure material. In the case of Figure 4c, the mean and deviation of the Gaussian DOS were also allowed to vary. For the temperature series in Figure 2 and 4, the same parameters from the corresponding 300K simulation were employed except the mean DOS energy and temperature. The exact values of the mean and deviation of the DOS for each dataset were summarised in Table S4, 5, 6, 7.